\documentclass[aps,pre,showpacs,singlecolumn]{revtex4}

\usepackage{bm}
\usepackage{graphicx}
\bibstyle{approve.bib}
\usepackage{amssymb}
\usepackage{amsmath}
\usepackage{esint}
\usepackage{epstopdf}
\usepackage{color}
\usepackage{gensymb}
\usepackage{mathtools}
\usepackage{suffix}

\newcommand{\be}{\begin{equation}}
\newcommand{\ee}{\end{equation}}
\newcommand{\bea}{\begin{eqnarray}}
\newcommand{\eea}{\end{eqnarray}}
\newcommand{\lan}{\left\langle}
\newcommand{\ran}{\right\rangle}
\newcommand{\br}{\mathbf{r}}

\newcommand{\bE}{\mathbf{E}}

\newcommand{\lm}{L_-}
\newcommand{\lp}{L_+}

\newcommand{\e}{\varepsilon}

\newcommand{\td}{\kappa d}
\newcommand{\ta}{\kappa a}

\newcommand{\ld}{\lambda_{\rm d}}
\newcommand{\lb}{\lambda_{\rm b}}

\newcommand{\ce}{{\rm c}}
\newcommand{\p}{{\rm p}}
\newcommand{\B}{{\rm B}}
\newcommand{\ex}{{\rm ex}}
\newcommand{\m}{{\rm m}}
\newcommand{\w}{{\rm w}}
\newcommand{\rb}{{\rm b}}

\begin{document}

\title{Theoretical modeling of polymer translocation: From the electrohydrodynamics of short polymers to the fluctuating long polymers}

\author{Sahin Buyukdagli$^{1}$\footnote{email:~\texttt{buyukdagli@fen.bilkent.edu.tr}}, Jalal Sarabadani$^{2,4}$\footnote{email:~\texttt{jalal@ipm.ir}}, and Tapio Ala-Nissila$^{3,4}$\footnote{email:~\texttt{Tapio.Ala-Nissila@aalto.fi}}}
\affiliation{$^{1}$Department of Physics, Bilkent University, Ankara 06800, Turkey\\
$^{2}$School of Nano Science, Institute for Research in Fundamental Sciences (IPM), Tehran 19395-5531, Iran\\
$^{3}$Department of Applied Physics and QTF Center of Excellence, Aalto University School of Science, P.O. Box 11000, FI-00076 Aalto, Espoo, Finland\\
$^{4}$Interdisciplinary Centre for Mathematical Modelling and Department of Mathematical Sciences, Loughborough University, Loughborough, Leicestershire LE11 3TU, United Kingdom}
\date{\small\it \today}

\begin{abstract}
The theoretical formulation of driven polymer translocation through nanopores is complicated by the combination of the pore electrohydrodynamics and the nonequilibrium polymer dynamics originating from the conformational polymer fluctuations. In this review, we discuss the modeling of polymer translocation in the distinct regimes of short and long polymers where these two effects decouple. For the case of short polymers where polymer fluctuations are negligible, we present a stiff polymer model including the details of the electrohydrodynamic  forces on the translocating molecule. We first show that the electrohydrodynamic theory can accurately characterize the hydrostatic pressure dependence of the polymer translocation velocity and  time in pressure-voltage-driven polymer trapping experiments. Then, we discuss the electrostatic correlation mechanisms responsible for the experimentally observed DNA mobility inversion by added multivalent cations in solid-state pores, and the rapid growth of polymer capture rates by added monovalent salt in $\alpha$-Hemolysin pores. In the opposite regime of long polymers where polymer fluctuations prevail, we review the iso-flux tension propagation (IFTP) theory which can characterize the translocation dynamics at the level of single segments. The IFTP theory is valid for a variety of polymer translocation and pulling scenarios. We discuss the predictions of the theory for fully flexible and rodlike pore-driven and end-pulled translocation scenarios, where exact analytic results can be derived for the scaling of the translocation time with chain length and driving force.
\end{abstract}

\date{\today}
\maketitle

\section{Introduction}

DNA is the key transmitter of the biological information carrying our genetic heritage. Fast and inexpensive access to this information is essential for various purposes ranging from the treatment of genetic diseases in medicine to the identification of harmful organisms in metagenomic sciences or DNA profiling in forensic sciences~\cite{rev1,rev2}. During the past three decades, this need has stimulated intensive research work on the development of efficient and low-cost biosequencing techniques such as the field-driven translocation of polymers through nanoscale pores~\cite{Tapsarev}. This biosensing approach consists of mapping the sequence of the polymer portion translocating through the pore from the current perturbations caused by the biopolymer~\cite{e1,e2,e3,e4,e5,e6,e7}. As the accuracy of this mapping depends sensitively on the duration of the current signal triggered by the presence of the translocating polymer, efficient use of this method requires a high degree of control on the dynamics of the molecule. At this point, one needs theoretical models able to predict the dependence of the polymer translocation dynamics on the experimentally controllable system parameters such as salt concentration, polymer charge and length, pore charge and size, and the external forces driving the translocation process.

The theoretical formulation of polymer translocation is a highly ambitious task. This complex transport process is indeed governed by a combination of effects such as pore electrohydrodynamics resulting from the electrophoretic (EP) and electroosmotic (EO) forces acting on the polymer, direct electrostatic polymer-membrane coupling, and entropic effects originating from conformational polymer fluctuations and steric polymer-membrane interactions. Two rather complementary approaches of distinct nature have so far been adopted for investigating polymer translocation: approaches based on coarse-grained conformational models and electrohydrodynamic formalisms. 

In the case of polymers longer than the translocated pores whose characteristic size $L_\m\sim10-100$ nm is comparable with the DNA persistence length, polymer fluctuations are substantial but the electrohydrodynamic forces can be assumed to act locally on DNA, i.e. exclusively on the polymer portion confined to the pore. This scale separation allows to bypass the details of the pore electrohydrodynamics that can be absorbed into the effective force $f$ driving the polymer and the effective pore friction $\eta_{\rm p}$ on it, enabling coarse-grained modeling of effects associated with non-equilibrium polymer conformations. Such coarse-grained models are easily amenable to molecular dynamics (MD) and Monte Carlo simulations~\cite{n1,n2,n3,n4,n5}, but even then it is a challenge to explicitly include electrostatic polymer-membrane interactions and they are usually assumed to be negligible.  On the theoretical side, a comprehensive theory for driven polymer translocation dynamics has been developed based on the idea of non-equilibrium tension propagation~\cite{ten1,ten2,ten3,jalal2014,jalal2017EPL,jalal2017SR}. The basic idea in this theory is to focus on the dynamics of a single degree of freedom, the translocation coordinate $s(t)$, and include all the many-body effects arising from the (non-equilibrium) chain conformations on the {\it cis} side of the membrane into a time-dependent friction $\eta_{\rm cis}(t)$. This leads into a Brownian dynamics type of equation for $s(t)$ which makes the problem both analytically and numerically tractable, and allows exact analytic results for the scaling of the translocation time as a function of the chain length. As explained in Sec. 3, this iso-flux tension propagation (IFTP) theory has been benchmarked for a variety of driven polymer translocation scenarios with excellent agreement with coarse-grained MD simulations and relevant experiments.

In the opposite regime of polymers whose length is comparable to the thickness of the translocated membrane, polymer fluctuations in the pore can be assumed to be negligible but the electrostatic polymer-pore interactions and the electrohydrodynamic pore effects have to be accurately taken into account.  A consistent electrohydrodynamic modeling of polymer translocation was initiated by Ghosal in Ref.~\cite{the1}. Via the coupled solution of the electrostatic Poisson-Boltzmann (PB) and hydrodynamic Navier-Stokes equations, Ghosal derived DNA translocation velocity as the superposition of the EP and EO velocity components. The role played by polymer-pore interactions on the unzipping of a DNA hairpin was investigated in Ref.~\cite{the2} without pore hydrodynamics. The effect of the EO flow on diffusion-limited polymer capture was studied in Ref.~\cite{the3} and the predictions of different electrostatic models~\cite{the4,the5} were compared with translocation experiments~\cite{e8}. Within a Smoluchowski formalism, we incorporated in Ref.~\cite{Buy2017} mean-field (MF) level electrostatic polymer-membrane interactions into the electrohydrodynamic model of Ref.~\cite{the1}. This unified polymer translocation theory was extended in Ref.~\cite{Buy2018I} to include electrostatic correlations. The extended theory was applied to the experiments of Ref.~\cite{e21} to explain the electrohydrodynamic mechanism behind the polyvalent-cation-induced DNA mobility reversal. In the same work, a new mechanism of facilitated polymer capture by charge-inverted EO flow was also identified. Finally, in Ref.~\cite{Buy2018II}, we revealed an electrostatic trapping mechanism enabling the extension of the polymer translocation time, which would allow to enhance the duration of the current readout in translocation experiments. Very recently, we have also taken a step towards a unified theory of polymer translocation by incorporating  the electrostatic coupling of the membrane with the \textit{cis} and \textit{trans} portions of the polymer outside the nanopore into the stiff polymer limit of the IFTP theory \cite{poly2018}.

In this article, we present a comparative review of the electrohydrodynamic and coarse-grained approaches described above. In the first part of the manuscript, we discuss in Sec.~\ref{elh} the electrohydrodynamic translocation model and its application to various experimental setups. First in Sec.~\ref{th}, we explain the theoretical framework of the approach. Section~\ref{resmf} is devoted to the application of the theory to pressure-voltage-driven translocation experiments in monovalent salt where the system is governed by MF electrohydrodynamics. Section~\ref{pol} is devoted to the translocation experiments with polyvalent salt where the high ion valency results in a departure from MF-level electrohydrodynamics. In Sec.~\ref{den}, we focus on polymer translocation through  $\alpha$-Hemolysin ($\alpha$HL) pores of subnanometer radius where the strong confinement results in polarization forces driving the system away from the MF transport regime. In the remaining part of the article, we focus on the regime of long coarse-grained polymers and review the iso-flux tension propagation theory able to account for the conformational polymer fluctuations during translocation. Our main results and prospects are discussed in the Summary and Conclusions section.

\section{Electrohydrodynamic approach to the translocation of short polymers}
\label{elh}

\begin{figure}
\begin{centering}
\includegraphics[width=.8\linewidth]{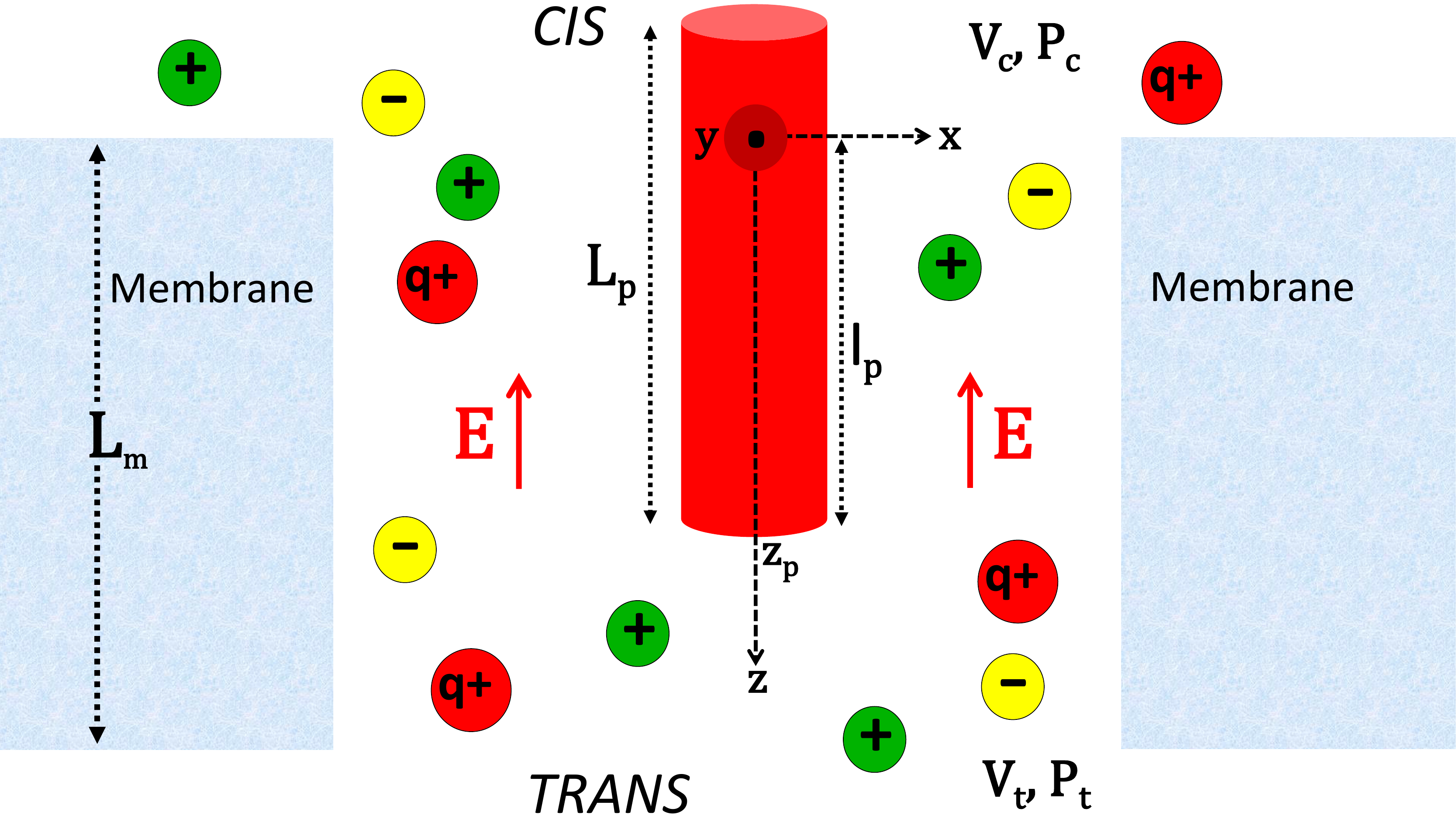}
\caption{(Color online) Schematic of a polymer translocating through a cylindrical pore of radius $d$, length $L_\m$, and negative surface charge density $-\sigma_\m$. The anionic polymer translocating on the $z$ axis is a cylinder of length $L_\p$, radius $a$, and surface charge density $-\sigma_\p$. $l_\p$ is the length of the polymer portion located in the pore. Translocation is driven by the externally applied voltage $\Delta V=V_{\rm t}-V_{\rm c}$ resulting in the electric field $\mathbf{E}=-\Delta V/L_\m\hat{u}_{\rm z}$, and the hydrostatic pressure $\Delta P=P_{\rm c}-P_{\rm t}$.}
\label{fig1}
\end{centering}
\end{figure}

In this section, we focus on the translocation of short polymers whose size is comparable to the size of the translocated nanopore. The comparable spatial scale of the polymer and the pore requires a detailed consideration of the pore electrohydrodynamics driving the translocation process. This point is the main motivation behind the electrohydrodynamic translocation model presented in this part. The configuration of the model is illustrated in Fig.~\ref{fig1}. A cylindrical pore of radius $d$ and total length $L_\m$ extends along the $z$ axis. The ends of the pore are in contact with an ion reservoir composed of $p$ ionic species, with the species $i$ of valency $q_i$ and bulk concentration $\rho_{{\rm b}i}$. The pore surface at $r=d$ carries a fixed negative charge distribution of density $\sigma_\m(\br)=-\sigma_\m\delta(r-d)$. The translocating polymer on the $z$ axis is modeled as a cylinder of radius $a$, total length $L_\p$, and surface charge density $\sigma_\p(\br)=-\sigma_\p\delta(r-a)$. Moreover, the polymer portion located in the pore has length $l_\p$. The reaction coordinate of the translocation is the location $z_\p$ of the lower end of the molecule. The translocation of the polymer from the cis side at $z=0$ to the trans side at $z=L_\m$ is induced by an externally applied hydrostatic pressure gradient  $\Delta P$ and electric voltage $\Delta V$.  In addition to these external driving forces, the polyelectrolyte is also subjected to direct polymer-membrane interactions characterized by the electrostatic potential $V_\p(z_\p)$. 

Section~\ref{th} reviews the electrohydrodynamic formalism of polymer translocation introduced in Refs.~\cite{Buy2017,Buy2018II}. In Sec.~\ref{resmf}, we present the application of this theory to solid-state pores and its comparison with pressure-voltage-driven translocation experiments~\cite{exp2}.  In Sec.~\ref{pol}, we discuss the effect of charge correlations on polymer translocation in polyvalent electrolytes and the resulting DNA mobility reversal~\cite{Buy2018II} observed in voltage-driven translocation experiments~\cite{e21}. Finally, in Sec.~\ref{den}, we investigate surface polarization effects on polymer translocation through $\alpha$HL  pores of subnanometer confinement.

\subsection{Theory}
\label{th}

\subsubsection{Electrohydrodynamic formalism of polymer translocation}

The translocation process is characterized by the polymer diffusion equation
\be\label{con}
\partial_tc(z_{\rm p},t)=-\partial_{z_{\rm p}}J(z_{\rm p},t)
\ee
where the polymer current is
\be
\label{cur}
J(z_{\rm p},t)=-D\partial_{z_{\rm p}}c(z_{\rm p},t)+v_{\rm p}(z_{\rm p})c(z_{\rm p},t).
\ee
In Eqs.~(\ref{con}) and~(\ref{cur}), the functions $c(z_{\rm p},t)$ and $J(z_{\rm p},t)$ stand respectively for the polymer density and flux. Moreover, the transverse diffusion coefficient of the cylindrical polymer is
$D=\ln(L_{\rm p}/2a)/(3\pi\eta L_{\rm p}\beta)$~\cite{cyl1}, with the inverse thermal energy $\beta=1/(k_{\rm B}T)$ and water viscosity $\eta=8.91\times10^{-4}$ Pa s. In Eq.~(\ref{cur}), the first and second terms correspond respectively to the diffusive flux component, and the convective flux component associated with the polymer velocity $v_{\rm p}(z_{\rm p})$. 

We restrict ourselves to the steady-state regime where the polymer flux becomes constant and uniform, i.e. $J(z_\p,t)=J_{\rm st}$. Introducing the effective polymer potential $U_\p(z_\p)$ defined by
\be\label{vp}
v_\p(z_\p)=-\beta DU'_\p(z_\p),
\ee
Eq.~(\ref{cur}) can be recast as
\be\label{jo}
J_{\rm st}=-De^{-\beta U_\p(z_\p)}\partial_{z_\p}\left[c(z_\p)e^{\beta U_\p(z_\p)}\right].
\ee
We integrate now Eq.~(\ref{jo}) with the absorbing boundary condition (BC) $c(z_\ex)=0$ at the pore exit  
\be
\label{ex}
z_\ex\equiv L_\p+L_\m,
\ee
and impose the polymer density on the cis side of the reservoir, i.e. $c(0)=c_{\rm cis}$. This yields 
\bea
\label{j1}
J_{\rm st}&=&\frac{Dc_{\rm cis}}{\int_0^{z_\ex}\mathrm{d}z\;e^{\beta \left[U_\p(z_\p)-U_\p(0)\right]}};\\
\label{c1}
c(z_\p)&=&c_{\rm cis}\frac{\int_{z_\p}^{z_\ex}\mathrm{d}z\;e^{\beta \left[U_\p(z)-U_\p(z_\p)\right]}}{\int_{0}^{z_\ex}\mathrm{d}z\;e^{\beta \left[U_\p(z)-U_\p(0)\right]}}.
\eea
The polymer population in the pore is given by the integral of Eq.~(\ref{c1}),
\be\label{N}
N=\int_0^{z_\ex}\mathrm{d}z_\p c(z_\p).
\ee
The polymer translocation frequency corresponding to the inverse translocation time is defined as the polymer flux per polymer population in the pore, i.e. $\tau_\p^{-1}=J_{\rm st}/N$. Moreover, the polymer capture rate is given by the polymer flux per reservoir concentration, i.e. $R_\ce=J_{\rm st}/c_{\rm cis}$. Using these definitions together with Eqs.~(\ref{j1})-(\ref{N}), the polymer capture rate and translocation time follow as
\bea
\label{rc}
R_\ce&=&\frac{D}{\int_0^{z_\ex}\mathrm{d}z\;e^{\beta \left[U_\p(z)-U_\p(0)\right]}};\\
\label{tauc}
\tau_\p&=&\frac{1}{D}\int_0^{z_\ex}\mathrm{d}z\;e^{-\beta U_\p(z')}\int_{z}^{z_\ex}\mathrm{d}z''e^{\beta U_\p(z'')}.
\eea

The rate $R_\ce$ corresponds to the average speed at which a successful polymer capture takes place. In the \textit{drift regime} characterized by weak polymer-pore interactions, the limit $V_p(z_p)\to0$ of Eq.~(\ref{rc}) yields
\be\label{34}
R_\ce=\frac{v_{dr}}{1-e^{-v_{dr}(L_m+L_p)/D}}\approx v_{dr},
\ee
where the second equality is valid for high voltages and a positive drift velocity. We finally note that for comparison with pressure-voltage trapping experiments,  the average translocation velocity will be also needed. The average polymer velocity is defined as
\be\label{bv0}
\lan v_\p\ran=\frac{\int_0^{z_{\rm ex}}\mathrm{d}z_\p c(z_\p)v_\p(z_\p)}{\int_0^{z_{\rm ex}}\mathrm{d}z_\p c(z_\p)}.
\ee

\subsubsection{Derivation of the  polymer velocity $v_\p(z_\p)$}

We first note that the evaluation of the polymer capture rate, translocation time, and average velocity defined in Eqs.~(\ref{rc})-(\ref{bv0}) requires the effective polymer potential $U_\p(z_\p)$. The calculation of this potential necessitates in turn the knowledge of the polymer velocity $v_\p(z_\p)$ in Eq.~(\ref{vp}). To derive the latter, we  first express the PB and Stokes equations for the net electrostatic potential $\phi(r)$ and liquid velocity $u_c(r)$  in the cylindrical nanopore,
\bea\label{pb}
&&\frac{1}{r}\partial_r\left[r\partial_r\phi(r)\right]+4\pi\ell_\B\left[\rho_\ce(r)+\sigma(r)\right]=0;\\
\label{st}
&&\frac{\eta}{r}\partial_r\left[r\partial_ru_c(r)\right]-e\rho_\ce(r)E+\frac{\Delta P}{L_\m}=0,
\eea 
where we introduced the radial distance $r$ from the pore axis, the Bjerrum length $\ell_\B=\beta e^2/(4\pi\e_{\rm w})$ with the solvent (water) permittivity $\e_{\rm w}=80$ and the unit charge $e$, and the density of mobile ions $\rho_\ce(r)$ and fix charges $\sigma(r)=-\sigma_\m\delta(r-d)-\sigma_\p\delta(r-a)$. Next, we eliminate from Eqs.~(\ref{pb}) and~(\ref{st}) the ion density $\rho_\ce(r)$, integrate the resulting equation, and impose the no-slip BC at the pore wall $u_\ce(d)=0$ and at the DNA surface $u_\ce(a)=v_\p(z_\p)$. We finally account for Gauss' law $\phi'(a)=4\pi\ell_\B\sigma_\p$ and also the force balance relation on the polymer $F_{\rm el}+F_{\rm dr}+F_{\rm b}=0$, with the electric force $F_{\rm el}=2\pi aL_\p eE$, the hydrodynamic drag force $F_{\rm dr}=2\pi aL_\p\eta u'_\ce(a)$, and the force $F_{\rm b}=-V'_\p(z_\p)$ associated with electrostatic polymer-membrane interactions. This yields the liquid and polymer velocities in the form
\bea\label{vels}
u_\ce(r)&=&\mu_{\rm e}E\left[\phi(d)-\phi(r)\right]-\beta D_\p(r)\frac{\partial V_\p(z_\p)}{\partial z_\p}+\frac{\Delta P}{4\eta L_\m}\left[d^2-r^2-2a^2\ln\left(\frac{d}{r}\right)\right];\\
\label{velp}
v_\p(z_\p)&=&v_{\rm dr}-\beta D_\p(a)\frac{\partial V_\p(z_\p)}{\partial z_\p},
\eea
with the effective polymer diffusion coefficient in the pore medium
\be\label{efd}
D_\p(r)=\frac{\ln(d/r)}{2\pi\eta L_\p\beta},
\ee
the coefficient of electrophoretic (EP) polymer mobility $\mu_{\rm e}=\e_{\rm w} k_{\rm B}T/(e\eta)$, and the drift velocity component induced by the external voltage and pressure,
\be
\label{vdr}
v_{\rm dr}=\frac{\mu_{\rm e}\Delta V}{L_\m}\left[\phi(d)-\phi(a)\right]+\frac{\gamma a^2\Delta P}{4\eta L_\m},
\ee
where we introduced the geometric factor
\be\label{gm}
\gamma=\frac{d^2}{a^2}-1-2\ln\left(\frac{d}{a}\right). 
\ee
The first term on the r.h.s. of the drift velocity Eq.~(\ref{vdr}) includes the effect of the voltage-induced EP force on DNA (the first term in the bracket) and the opposing force from the electroosmotic (EO) flow drag (the second term in the bracket). The second term of Eq.~(\ref{vdr}) corresponds in turn to the contribution from the pressure-induced streaming flow to the DNA velocity. Then, the second term of Eq.~(\ref{velp}) brings the effect of electrostatic polymer-membrane interactions on the polymer velocity. As a result of the no-slip relation $v_\p(z_\p)=u_\ce(a)$, the terms on the r.h.s. of Eq.~(\ref{vels}) clearly indicate the contribution from the same effects to the convective liquid velocity $u_\ce(r)$. Integrating now Eq.~(\ref{vp}) with Eq.~(\ref{velp}), one finally obtains the effective polymer potential in Eqs.~(\ref{rc})-(\ref{bv0}) as
\be
\label{polp}
U_\p(z_\p)=\frac{D_\p(a)}{D}V_\p(z_\p)-\frac{v_{\rm dr}}{\beta D}z_\p.
\ee

\subsubsection{Derivation of the interaction potential $V_\p(z_\p)$}

We explain next the derivation of the electrostatic polymer-membrane interaction potential $V_\p(z_\p)$ in the MF regime of weak surface charges and physiological monovalent salt concentrations.  The extension of this calculation beyond MF electrostatics is rather involved and this generalization can be found in Refs.~\cite{Buy2018I,Buy2018II}. In the MF linear response regime, the polymer-membrane interaction potential induced by the electrostatic coupling between the membrane potential  and the polymer charges $Q_{\rm pol}=2\pi a l_\p\sigma_\p$ located in the pore reads
\be\label{Vp}
V_\p(z_\p)=-2\pi a\sigma_\p k_\B T\phi_\m(a) l_\p(z_\p).
\ee
In Eq.~(\ref{Vp}), the potential $\phi_\m(r)$ induced solely by the membrane charges is obtained from the solution of the PB Eq.~(\ref{pb}) without the polymer charge, i.e.
\be
\label{phim}
\phi_\m(r)=\lim_{\sigma_\p\to0}\phi(r).
\ee
The calculation of the potentials $\phi(r)$ and $\phi_\m(r)$ can be found in Ref.~\cite{Buy2017}. Moreover, the position-dependent length of the polymer portion in the pore reads
\bea
\label{lpzp}
l_\p(z_\p)&=&z_\p\theta(L_--z_\p)+L_-\theta(z_\p-L_-)\theta(L_+-z_\p)+(z_{\rm ex}-z_\p)\theta(z_\p-L_+),
\eea
where we introduced the auxiliary lengths 
\be\label{lpm}
L_-=\mathrm{min}(L_\m,L_\p);\hspace{5mm}L_+=\mathrm{max}(L_\m,L_\p). 
\ee
The terms on the r.h.s. of Eq.~(\ref{lpzp}) are associated with the regimes of polymer capture ($z_\p<L_-$), translocation at constant length ($L_-<z_\p<L_+$), and polymer escape ($z_\p>L_+$), respectively. Finally, defining the characteristic inverse lengths associated with the drift velocity in Eq.~(\ref{vdr}) and the electrostatic interaction potential in Eq.~(\ref{Vp}),
\be
\label{25}
\ld=\frac{v_{dr}}{D};\hspace{5mm}\lb=-2\pi a\sigma_p\phi_m(a)\frac{D_p(a)}{D},
\ee
the polymer translocation velocity in Eq.~(\ref{velp}) and the interaction potential in Eq.~(\ref{polp}) take the simpler forms
\bea\label{vpz}
v_\p(z_\p)&=&v_{\rm dr}-D\lambda_b\left[\theta(L_--z_\p)-\theta(z_\p-L_+)\right];\\
\label{up2}
\beta U_\p(z_\p)&=&\lb l_\p(z_\p)-\ld z_\p.
\eea

The inverse lengths $\lambda_{\rm d,b}$ in Eq.~(\ref{25}) allow to characterize the polymer capture and translocation dynamics  within the \textit{drift-driven} regime $\ld\gg\lb$ of weak polymer-membrane interactions and the \textit{barrier-driven} regime $\lb\gg\ld$  where the electrostatic polymer-membrane coupling takes over the voltage and/or pressure-induced drift force~\cite{Buy2018II}. The explicit form of Eqs.~(\ref{rc})-(\ref{bv0}) obtained with the velocity in Eq.~(\ref{vpz}) and the potential profile in Eq.~(\ref{up2}) are given in Appendix~\ref{ap1} in terms of the inverse lengths defined in Eq.~(\ref{25}).  Evaluating now the average polymer velocity of Eq.~(\ref{bv0}) with Eqs.~(\ref{vpz}) and~(\ref{up2}), one gets
\bea\label{vp2}
\lan v_\p\ran=D\ld-D\lambda_b\frac{J_1-J_3}{J_1+J_2+J_3},
\eea
where the coefficients $J_i$ are also given in Appendix~\ref{ap1}. Finally, the translocation time in Eq.~(\ref{tauc}) follows as
\be\label{tauc2}
\tau_\p=\tau_1+\tau_2+\tau_3,
\ee
where the explicit forms of the characteristic time for polymer capture $\tau_1$, translocation $\tau_2$, and escape $\tau_3$ are given in Appendix~\ref{ap2}.

\subsection{Polymer conductivity of solid-state pores: MF electrohydrodynamics with monovalent salt}
\label{resmf}

We consider polymer translocation events in solid-state pores and monovalent salt solutions where the electrohydrodynamic interactions are characterized by MF electrostatics.   The DNA surface charge density is fixed to the value $\sigma_\p=0.4$ $e/\mbox{nm}^2$ previously obtained in Ref.~\cite{the12} by fitting experimental current blockade data.  In Ref.~\cite{Buy2018II}, it was shown that both in the barrier-driven regime $\lb\gg\ld$ and drift-dominated regime $\ld\gg\lb$, the translocation velocity in Eq.~(\ref{vp2}) can be well approximated by $\lan v_\p\ran\approx D(\ld-\lb)$. Passing to the Debye-H\"{u}ckel (DH) limit of strong salt, this approximation yields~\cite{Buy2018II}
\be\label{bv}
\lan v_\p\ran\approx\frac{(f_\p\sigma_\p-f_\m\sigma_\m)}{g\kappa\eta}\frac{e\Delta V}{L_\m}+\frac{\gamma a^2\Delta P}{4\eta L_\m}-\frac{e^2\sigma_\p\sigma_\m\ln(d/a)}{g\eta\e_{\rm w}\kappa^2L_\p},
\ee
where we used the DH screening parameter $\kappa^2=8\pi\ell_\B\rho_\rb$ and introduced the geometric coefficients $g$ and $f_{\m,\p}$ given in Appendix~\ref{ap1}. In Eq.~(\ref{bv}), the first term on the r.h.s. takes into account the electrophoretic (EP) drift force by the electric field on the polymer charges (positive term) and the electroosmotic (EO) flow drag induced by the counterions attracted by the charged membrane (negative term). In the present case of anionic polymers translocating through like-charged membranes, the EO flow opposing the EP drift reduces the polymer velocity, i.e. $\sigma_\m\uparrow R_\ce\downarrow\lan v_\p\ran\downarrow$. Then, the second term in Eq.~(\ref{bv}) corresponds to the force exerted on the polymer by the pressure-induced streaming flow through the pore. Finally, the third term including the product $\sigma_\p\sigma_\m>0$ accounts for the electrostatic polymer-membrane interactions. Due to the resulting like-charge repulsion, the negative interaction term acts as an electrostatic barrier hindering the polymer capture by the nanopore.

\begin{figure}
\begin{centering}
\includegraphics[width=1\linewidth]{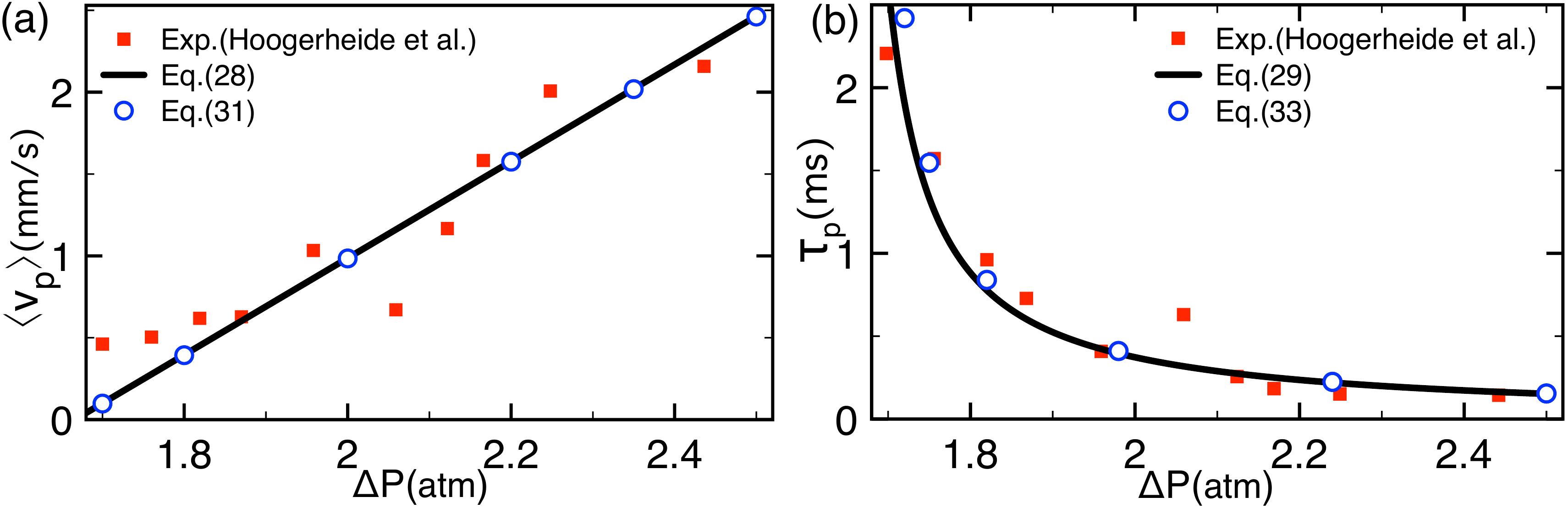}
\caption{(Color online) (a) Pressure dependence of the average translocation velocity $\lan v_\p\ran$ obtained from Eq.~(\ref{vp2}) (solid curve) and the drift formula~(\ref{vm}) (circles). (b) Translocation time $\tau_\p$ from Eq.~(\ref{tauc2}) (solid curve) and the drift Eq.~(\ref{tausc}). In (a), the experimental polymer velocity data was taken from Fig. S3 of the supporting information of Ref.~\cite{exp2}. The average escape time data in (b) is from Fig. 4(b) of Ref.~\cite{exp2}. The numerical values of the model parameters are given in the main text. The results are from Ref.~\cite{Buy2018II}.}
\label{fig2}
\end{centering}
\end{figure}

\subsubsection{Comparison with pressure-voltage trapping experiments}

Equation~(\ref{bv}) shows that in the drift-driven regime, the average velocity rises linearly with the external voltage $\Delta V$. This linear dependence has been observed in experiments and simulations~\cite{e14,sim}. To understand the pressure dependence of the translocation velocity and time, we now focus on the pressure-driven translocation experiments. Figure~\ref{fig2}(a) compares the average polymer velocity in Eq.~(\ref{vp2}) (solid curve) with the experimental data of Ref.~\cite{exp2} (squares). The numerical values of the model parameters taken from Ref.~\cite{exp2} are the negative external voltage $\Delta V=-100$ mV opposing the drag of the streaming flow, the electrolyte concentration $\rho_\rb=1.6$ M, the number of monomers in the DNA sequence $N=615$ bps corresponding to the DNA length $L_\p=180$ nm, and the pore radius $d=5$ nm. The values of the membrane thickness $L_\m=200$ nm and charge $\sigma_\m=0.13$ $e/\mbox{nm}^2$ were adjusted to obtain the optimal agreement with the magnitude of the translocation velocity data. 

Figure~\ref{fig2}(a) shows that within the experimental scattering, the theoretical result agrees well with the polymer velocity data. For an analytical insight into the pressure dependence of the experimental data, we recast Eq.~(\ref{bv}) in the form
\be\label{vm}
\lan v_\p\ran\approx\frac{\gamma a^2}{4\eta L_\m}\left(\Delta P-\Delta P^*\right),
\ee
with the critical pressure where the translocation velocity vanishes and the polymer gets trapped
\be
\label{prcr}
\Delta P^*=-\frac{4\left(f_\p\sigma_\p-f_\m\sigma_\m\right)}{\gamma ga^2\kappa}e\Delta V +\frac{4\ln(d/a)e^2\sigma_\p\sigma_\m L_\m}{\gamma ga^2\e_\w\kappa^2L_\p}.
\ee
Equation~(\ref{vm}) reported in Fig.~\ref{fig2} (a) by circles indicates that the average polymer velocity grows linearly with the pressure gradient.  

Figure~\ref{fig2}(b) compares the theoretical translocation time $\tau_\p$ in Eq.~(\ref{tauc2}) (solid curve) with the experimental escape times of Ref.~\cite{exp2}. The theoretical result obtained with the same parameters as in Fig.~\ref{fig2}(a) can accurately reproduce the general trend of the experimental data. To identify the scaling of the experimental time data with the pressure gradient, we note that the occurrence of a successful polymer translocation necessitates the polymer of average velocity $\lan v_\p\ran$ to travel the total distance $z_{\rm ex}=L_\p+L_\m$ over the time $\tau_\p$. This allows to approximate the translocation time as $\tau_\p\approx z_{\rm ex}/\lan v_\p\ran$. Using Eq.~(\ref{vm}), this yields
\be\label{tausc}
\tau_\p\approx\frac{4\eta L_\m\left(L_\p+L_\m\right)}{\gamma a^2\left(\Delta P-\Delta P^*\right)}.
\ee
The approximative formula~(\ref{tausc}) reported in Fig.~\ref{fig2}(b) by open circles indicates that the pronounced rise of the translocation time at low pressures is characterized by the inverse power law scaling $\tau_\p\sim\left(\Delta P-\Delta P^*\right)^{-1}$.

\subsubsection{Salt and polymer length dependence of pressure-voltage-driven translocation events}

\begin{figure}
\begin{centering}
\includegraphics[width=1.\linewidth]{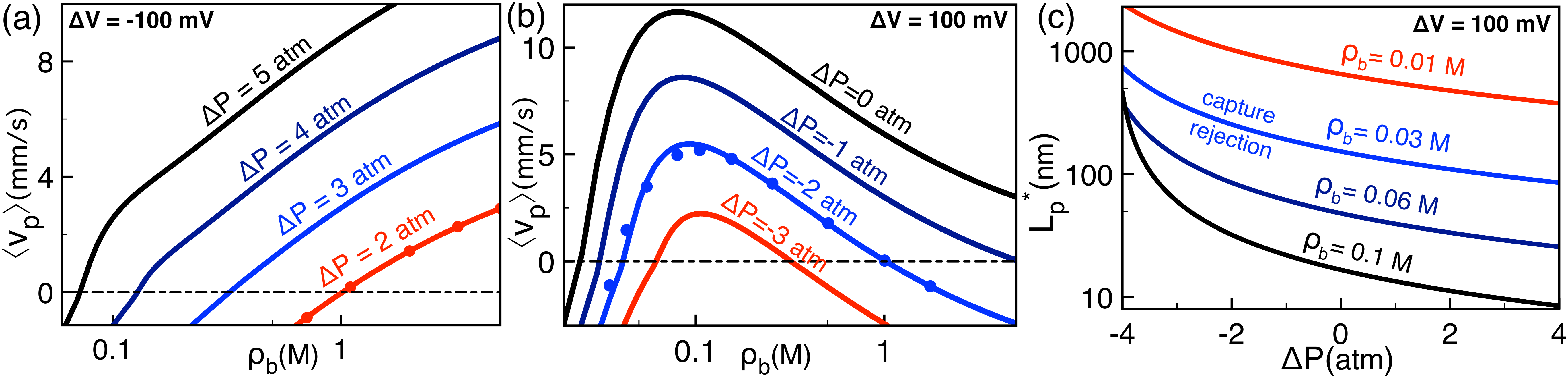}
\caption{(Color online)  Average polymer velocity from Eq.~(\ref{vp2}) (solid curves) and the drift formula~(\ref{vm}) (dots) versus salt concentration in (a) voltage-limited ($\Delta V=-100$ mV) and (b) voltage-driven translocation ($\Delta V=100$ mV). (c) Characteristic polymer length in Eq.~(\ref{lcr}) separating the polymer capture and rejection regimes versus the pressure $\Delta P$ in voltage-driven translocation with $\Delta V=100$ mV. The remaining parameters are the same as in Fig.~\ref{fig2}. The results are from Ref.~\cite{Buy2018II}.}
\label{fig3}
\end{centering}
\end{figure}

We discuss here the influence of the ion density and polymer length on pressure-voltage-driven polymer translocation events. Figures~\ref{fig3}(a) and (b) display the ion density dependence of the polymer translocation velocity obtained from Eq.~(\ref{vp2}) (solid curves) and the drift formula~(\ref{bv}) (dots at $\Delta P=\pm2$ atm). We first focus on Fig.~\ref{fig3}(a) where polymer translocation is driven by the streaming flow ($\Delta P>0$) and limited by voltage ($\Delta V<0$).  In this case, one notes that added salt increases the translocation velocity ($\rho_\rb\uparrow\lan v_\p\ran\uparrow$) and turns the velocity from negative to positive. For an analytical insight into the enhancement of polymer capture by added salt, we Taylor-expand Eq.~(\ref{bv}) in the high salt density regime $\kappa a\gg1$ and $\kappa d\gg1$. This yields
\be\label{dil1}
\lan v_\p\ran\approx\frac{(\sigma_\p-\sigma_\m)e\Delta V}{\eta L_\m\kappa}+\frac{\gamma a^2\Delta P}{4\eta L_\m}.
\ee 
The screening parameter $\kappa$ in the first term of Eq.~(\ref{dil1}) indicates that the salt-induced growth of the polymer velocity in Fig.~\ref{fig3}(a) results from the enhanced screening of the EP drift force opposing the polymer capture. 

In the opposite case of voltage-pressure driven ($\Delta V>0$) and pressure-limited translocation ($\Delta P<0$) displayed in Fig.~\ref{fig3}(b), the salt dependence of the translocation velocity is non-monotonic. More precisely, in the strong salt regime $\rho_\rb>0.1$ M, the salt-screening of the EP polymer mobility in Eq.~(\ref{dil1}) is seen to reduce the translocation velocity ($\rho_\rb\uparrow\lan v_\p\ran\downarrow$) and switch its sign from positive to negative. The characteristic salt density for polymer trapping follows from Eq.~(\ref{dil1}) as
\be
\label{dil2}
\rho_>^*\approx\frac{2}{\pi\ell_\B}\left[\frac{(\sigma_\p-\sigma_\m)e\Delta V}{\gamma a^2\Delta P}\right]^2.
\ee
In accordance with Fig.~\ref{fig3}(a) and (b), Eq.~(\ref{dil2}) predicts a drop of the trapping density with an increasing magnitude of the pressure gradient, $|\Delta P|\uparrow\rho_>^*\downarrow$. 

We focus now on the dilute salt regime $\rho_\rb<0.1$ M of Fig.~\ref{fig3}(b) where one notes the enhancement of the translocation velocity with added salt $\rho_\rb\uparrow\lan v_\p\ran\uparrow$ and the presence of a second characteristic salt density where the velocity vanishes and the polymer gets trapped. For an insight into these features, we Taylor-expand Eq.~(\ref{bv}) in the dilute salt regime $\kappa a\ll1$ and $\kappa d\ll1$ to obtain
\bea
\label{dil3}
\lan v_\p\ran&\approx&\frac{(a_\p\sigma_\p-a_\m\sigma_\m)e\Delta V}{\eta L_\m}+\frac{\gamma a^2\Delta P}{4\eta L_\m}-\frac{da\ln(d/a)}{d^2-a^2}\frac{k_\B T\sigma_\p\sigma_\m}{\eta L_\p\rho_\rb}, 
\eea
where the expansion coefficients are introduced as
\be\label{apm}
a_\p=-\frac{a}{2}+\frac{ad^2\ln(d/a)}{d^2-a^2};\hspace{3mm}a_\m=\frac{d}{2}-\frac{a^2d\ln(d/a)}{d^2-a^2}.
\ee
Equation~(\ref{dil3}) shows that in the dilute salt regime of Fig.~\ref{fig3}(b), the negative translocation velocity corresponding to the polymer rejection regime is induced by repulsive polymer-membrane interactions (the third term on the r.h.s.). The screening of these interactions by added dilute salt results in the rise of the translocation velocity ($\rho_\rb\uparrow\lan v_\p\ran\uparrow$) and the reversal of its sign from negative to positive. The characteristic dilute salt density for polymer trapping follows from Eq.~(\ref{dil3}) as
\be\label{dil4}
\rho_<^*\approx\frac{4da\ln(d/a)L_\m}{(d^2-a^2)L_\p}\frac{k_\B T\sigma_\p\sigma_\m}{\gamma a^2\Delta P+4(a_\p\sigma_\p-a_\m\sigma_\m)e\Delta V}.
\ee
In agreement with Fig.~\ref{fig3}(b), Eq.~(\ref{dil4}) predicts the increase of the lower characteristic salt density with increasing magnitude of the negative pressure, i.e. $|\Delta P|\uparrow \rho_>^*\uparrow$.

Finally, we consider the effect of the polymer length on the  translocation dynamics. According to Eq.~(\ref{bv}), the reduction of the polymer length enhances the repulsive barrier term and reduces the polymer velocity, i.e. $L_\p \downarrow \lan v_\p\ran\downarrow$. The slowing down of the translocation by finite polymer length results from the competition between the externally applied drift force and repulsive polymer-pore interactions; the pressure-voltage-induced drift acts on the entire polymer of length $L_p$  while the electrostatic barrier originates solely from the polymer portion enclosed by the pore. As a result, the net drag force on the translocating polymer drops with the length of the molecule. Due to this balance, the polymer velocity in Eq.~(\ref{bv}) decreases inversely proportional to the polymer length
\be\label{l1}
\lan v_\p\ran\approx v_{\rm dr}\left(1-\frac{L^*_\p}{L_\p}\right),
\ee
with the critical molecular length for polymer trapping
\be\label{lcr}
L_\p^*=\frac{4e^2\sigma_\p\sigma_\m\ln(d/a)L_\m}{\gamma a^2\e_\w g\kappa^2\Delta P+4\e_\w\kappa\left(f_\p\sigma_\p-f_\m\sigma_\m\right)e\Delta V}.
\ee
The characteristic length $L_\p^*$ is plotted in Fig.~\ref{fig3}(c).  First, one notes that the competition between the streaming current and the repulsive barrier leads to the decay of the characteristic length in Eq.~(\ref{lcr}) with pressure, i.e. $\Delta P\uparrow L_\p^*\downarrow$. As illustrated in the same figure, the dilute salt expansion of Eq.~(\ref{lcr}) 
\be\label{lcr2}
L_\p^*\approx\frac{4da\ln(d/a)L_\m}{(d^2-a^2)\rho_\rb}\frac{k_BT\sigma_\p\sigma_\m}{\gamma a^2\Delta P+4(a_\p\sigma_\p-a_\m\sigma_\m)e\Delta V},
\ee
indicates that due to the same balance between the drift force and the electrostatic barrier, the critical length also drops with added salt, i.e. $\rho_\rb\uparrow L_\p^*\downarrow$. In the next section, we investigate the deviation from the MF polymer transport behavior studied herein by added polyvalent cations.

\subsection{Correlation-induced DNA mobility inversion by polyvalent counterions in solid-state pores}
\label{pol}

In this section, we reconsider the polyvalent cation-induced DNA mobility inversion observed by the experiments of Ref.~\cite{e21} and theoretically investigated in Ref.~\cite{Buy2018I}. The multivalency of counterions requires the inclusion of charge correlations to the MF potential obtained from Eq.~(\ref{pb}). The details of this correlation-corrected scheme can be found in Ref.~\cite{Buy2018I}.

\begin{figure}
\begin{centering}
\includegraphics[width=.6\linewidth]{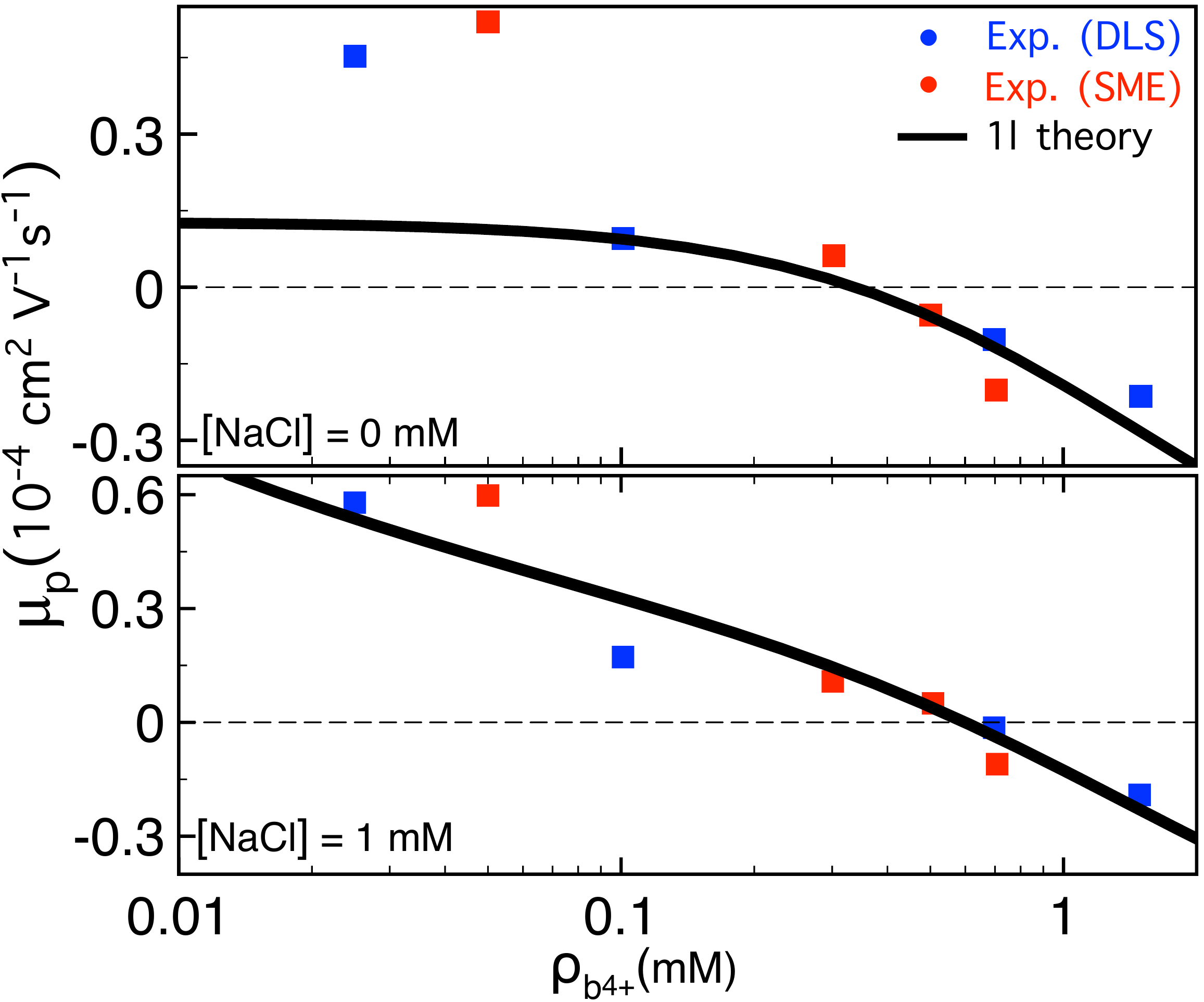}
\caption{(Color online) Electrophoretic DNA mobility $\mu_\p=v_{\rm dr}/E$ against $\mbox{Spm}^{4+}$ concentration in the electrolyte mixture NaCl$+\mbox{SpmCl}_4$. The monovalent cation density is $\rho_{\rb+}=0$ mM (top) and $1$ mM (bottom). Solid curves:  Theoretical prediction of Eq.~(\ref{vdr}). Symbols: dynamic light scattering (DLS) and single molecule electrophoresis (SME) data of Ref.~\cite{e21}. ds-DNA molecule has radius $a=1$ nm and effective surface charge density $\sigma_\p=-0.12$ $e/\mbox{nm}^2$. The nanopore has radius $d=10$ nm and fixed surface charge density $\sigma_\m=-0.006$ $e/\mbox{nm}^2$. The results are from Ref.~\cite{Buy2018I}.}
\label{fig6I}
\end{centering}
\end{figure}

Figure~\ref{fig6I} illustrates the polymer mobility $\mu_\p=v_{\rm dr}/E$ against the concentration of quadrivalent spermine ($\mbox{Spm}^{4+}$)  molecules in the NaCl$+\mbox{SpmCl}_4$ solution of two different NaCl density. The plots compare the theoretical result $=\mu_\p=\mu_{\rm e}\left[\phi(d)-\phi(a)\right]$ obtained from Eq.~(\ref{vdr}) (solid curves) with the experimental dynamic light scattering (DLS) and single molecule electrophoresis (SME) experiments of Ref.~\cite{e21} (squares).  The nanopore and polymer charge densities given in the caption are the free parameters of the model that were adjusted to give the best agreement with the experimental data. The pore radius was in turn fixed to the value $d=10$ nm corresponding to the characteristic radial size of solid-state nanopores.

As expected from MF electrophoresis, the dilute $\mbox{Spm}^{4+}$ regime of Fig.~\ref{fig6I} is characterized by a positive DNA mobility  $\mu_\p>0$  corresponding to the drift of the negatively charged polymer oppositely to the external electric field $\bE$, i.e. from the \textit{cis} to the \textit{trans} side of the membrane (see Fig.~\ref{fig1}).  However, the increment of the $\mbox{Spm}^{4+}$ concentration reduces the DNA mobility and switches its sign from positive to negative, indicating the reversal of the DNA translocation velocity  from the \textit{cis}-\textit{trans} to the \textit{trans}-\textit{cis} direction.  This corresponds to a non-MF charge transport picture where the anionic molecule moves parallel with the applied field $\bE$. Moreover, the comparison of the top and bottom figures indicates that added monovalent salt weakens charge correlations and rises the DNA mobility ($\rho_{\rb+}\uparrow\mu_\p\uparrow$) and the characteristic spermine density $\rho_{\rb4+}^*$ for mobility inversion ($\rho_{\rb+}\uparrow\rho^*_{\rb4+}\uparrow$). Within the experimental scattering, the theory can account for these features with reasonable quantitative accuracy.

\begin{figure}
\begin{centering}
\includegraphics[width=0.7\linewidth]{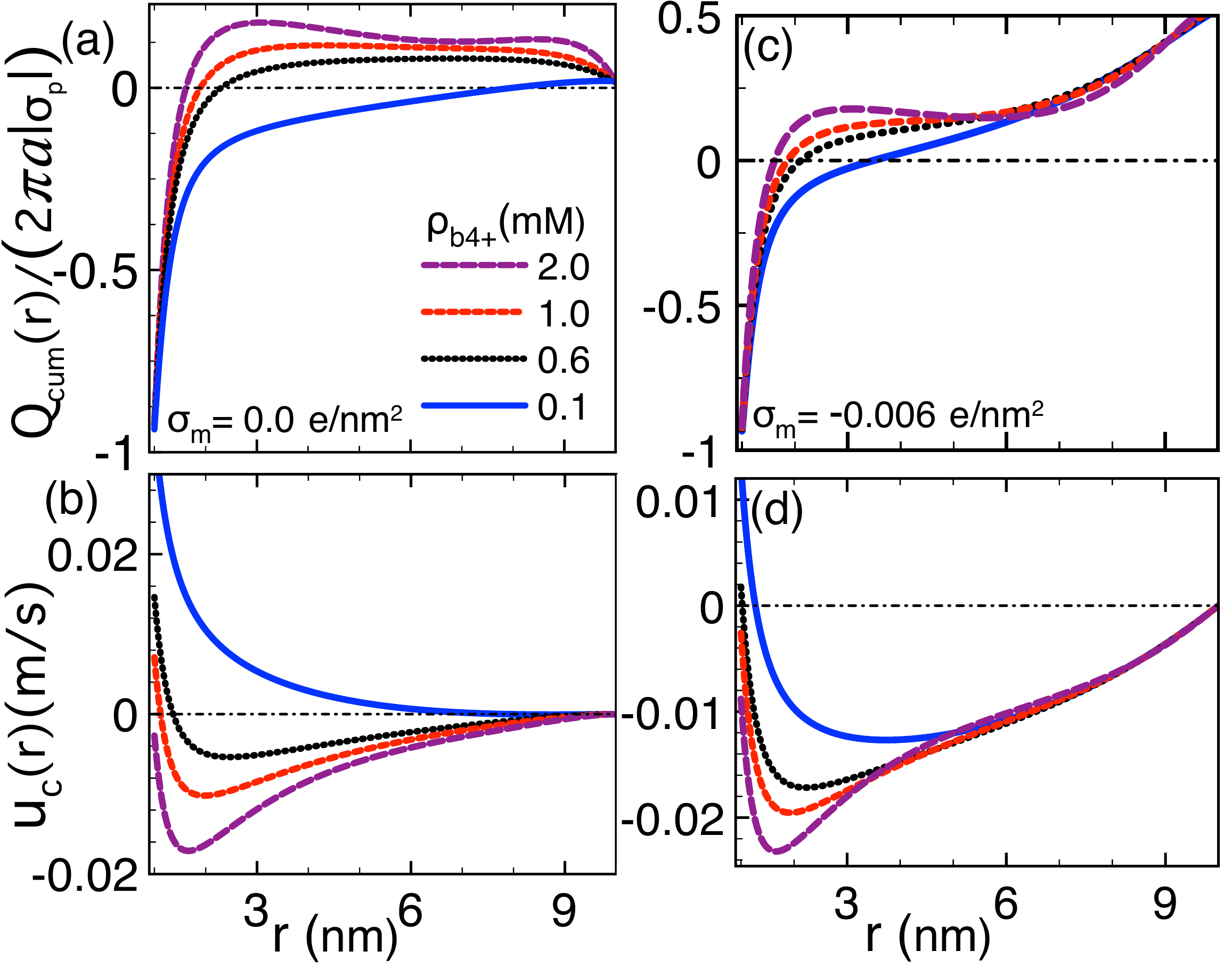}
\caption{(Color online) Rescaled cumulative charge density $Q_{\rm cum}(r)/(2\pi a|\sigma_\p|)$ (top plots) and electrolyte velocity $u_\ce(r)$ (bottom plots) in (a-b) neutral and (c-d) weakly charged nanopores with fixed surface charge density $\sigma_\m=-0.006$ $e/\mbox{nm}^2$. The applied voltage is $\Delta V=120$ mV, the nanopore length $L_\m=34$ nm, and the monovalent counterion concentration $\rho_{\rb+}=1$ mM. The other parameters are the same as in Fig.~\ref{fig6I}. The results are from Ref.~\cite{Buy2018I}.}
\label{fig7I}
\end{centering}
\end{figure}

The electrohydrodynamic mechanism driving the DNA velocity reversal is illustrated in Fig.~\ref{fig7I} where we plot the cumulative charge density (top plots) defined as
\be\label{cum}
Q_{\rm cum}(r)=2\pi\int_a^r\mathrm{d}r'r'\left[\rho_\ce(r')+\sigma_\p(r')\right],
\ee
and the convective liquid velocity profile obtained from Eq.~(\ref{vels}) (bottom plots). Equation~(\ref{cum}) corresponds to the net charge of the DNA and its counterions.  $\rho_\ce(r)$ is the local mobile charge density of the PB Eq.~(\ref{pb}). In order to emphasize first the role played by electrophoresis only, in Figs.~\ref{fig7I}(a) and (b), the EO flow was switched off by considering a neutral membrane ($\sigma_\m=0$). In the dilute spermine regime with $\rho_{\rb4+}=0.1$ mM (blue curve in Fig.~\ref{fig7I}(a)), as one approaches the pore wall from the DNA surface, the gradual screening of the DNA charges by the counterions leads to a decreasingly negative total charge density $Q_{\rm cum}(r)\leq0$. This net negative charge coupled to the external field $\bE$ results in the motion of the DNA and its counterions along the positive $z$ axis (see Fig.~\ref{fig1}), i.e. $u_\ce(r)\geq0$ and $v_{\rm dr}=u_\ce(a)>0$ (blue curve in Fig.~\ref{fig7I}(b)).

In the larger spermine concentration regime $\rho_{\rb4+}=0.6$ mM (black curves) and $1.0$ mM (red curves), beyond the characteristic distance of $\sim1$ nm from the DNA surface, electrostatic correlations enhanced by the multivalency of $\mbox{Spm}^{4+}$ molecules switch the cumulative charge from negative to positive, indicating the occurrence of DNA charge inversion (CI). Consequently, in the vicinity of the DNA molecule, the charged liquid changes its direction and flows parallel  with the external field $\bE$, i.e. $u_\ce(r)<0$. One however notes that at those $\mbox{Spm}^{4+}$ densities where CI is not strong enough to invert the electrophoretic force on DNA, the molecule continues to translocate opposite to the external electric field $\bE$, i.e. $v_{\rm dr}=u_\ce(a)>0$.  Upon further increase of the $\mbox{Spm}^{4+}$ concentration to the critical value $\rho_{\rb4+}^*=2.0$ mM (purple curves), stronger charge correlations amplify the inverted charge density. As a result, the hydrodynamic drag by the charge inverted liquid on the DNA surface takes over the electric force on the DNA charges, resulting in the reversal of the DNA velocity from positive to negative ($v_{\rm dr}<0$) and the direction of the molecule from the \textit{trans} to the \textit{cis} side.

These results show that the DNA mobility reversal is driven by a strong enough DNA charge inversion. The additional effect of the EO flow drag on this peculiarity is displayed in Figs.~\ref{fig7I}(c) and (d) where we included the finite membrane charge density in Fig.~\ref{fig6I}. The comparison of the left and right plots shows that the attraction of counterions by the anionic membrane charge amplifies the positive liquid density $Q_{\rm cum}(r)$. As the corresponding EO flow positively adds to the hydrodynamic drag force exerted by the charge inverted liquid, the characteristic spermine density for DNA velocity reversal drops with the membrane charge,  i.e. $|\sigma_m|\uparrow\rho_{b4+}^*\downarrow$. Next, we study the transport properties of biological nanopores where the strong pore confinement results in pronounced correlation effects even in monovalent electrolytes.

\subsection{Polymer conductivity of biological pores: image-charge barrier against drift force}
\label{den}

In this part, we investigate the voltage-driven polymer transport properties of $\alpha$HL channels where the strong pore confinement and the low membrane permittivity $\e_\m\approx2\ll\e_\w$ gives rise to an image-charge barrier opposing the drift force on the polymer.  The nanopore of radius $d=8.5$ {\AA} and length $L_\m=5$ nm contains a monovalent KCl salt with bulk concentration $\rho_{\pm \rb}=\rho_\rb$. The pore confines as well a ss-DNA molecule with radius $a\approx5$  {\AA} and linear charge density  $\tau\equiv-2\pi a\sigma_\p=0.29$ e/{\AA}. We first focus on the experimentally observed rapid rise of polymer translocation rates with added salt. Within a phenomenological approach, this effect was explained in Ref.~\cite{e4} by image-charge interactions. Within our translocation model, we intend to bring an analytical explanation to this peculiarity. 

\begin{figure}
\begin{centering}
\includegraphics[width=1.0\linewidth]{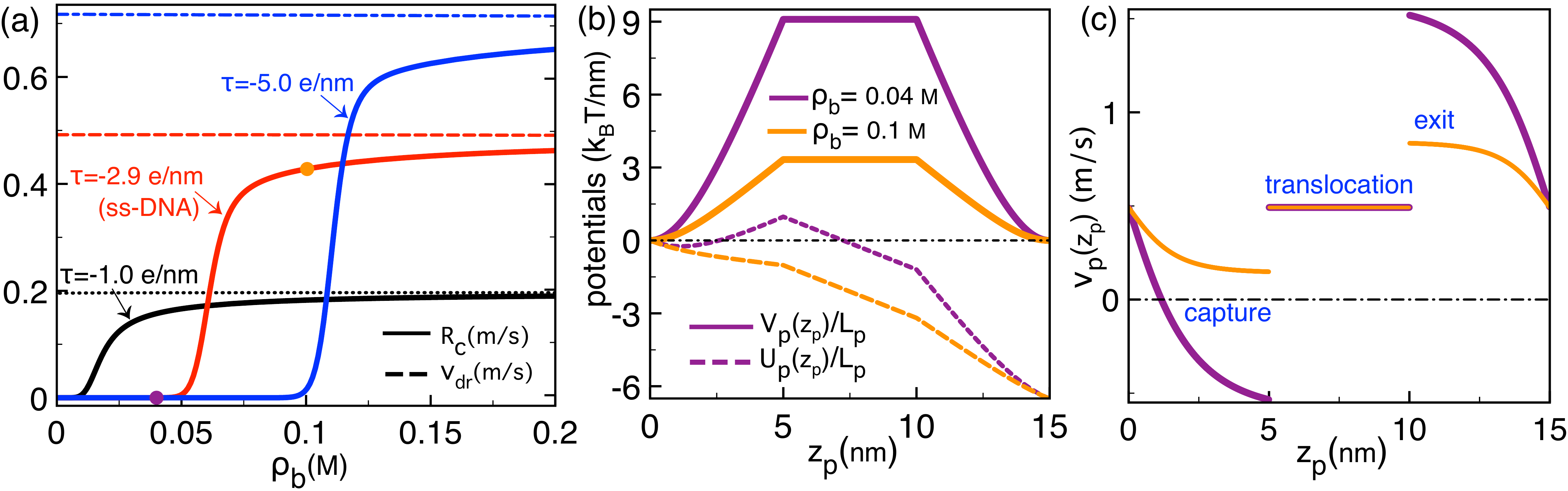}
\caption{(Color online) (a) Polymer capture rate $R_\ce$ in Eq.~(\ref{rc}) (solid curves) and drift velocity $v_{\rm dr}$ of Eq.~(\ref{vdr}) (dashed curves) in $\alpha$HL pores against the bulk salt concentration $\rho_\rb$ at various linear polymer charge density values $\tau=-2\pi a\sigma_\p$. (b) Polymer-pore interaction potential $V_\p(z_\p)$ including image-charge forces (solid curves), effective potential $U_\p(z_\p)$ from Eq.~(\ref{polp}) (dashed curves), and (c) velocity profile $v_\p(z_\p)$ of Eq.~(\ref{velp}) at the salt densities $\rho_\rb=0.04$ M (purple) and $0.1$ M (orange). In all plots, the pore and polymer lengths and radii are $L_\m=5$ nm, $L_\p=10$ nm, $d=8.5$ {\AA}, and $a=5$ {\AA}. The membrane is overall neutral ($\sigma_\m=0$) and the applied voltage is $\Delta V=120$ mV.}
\label{fig4}
\end{centering}
\end{figure}

$\alpha$HL pores are characterized by a non-uniform surface charge distribution with alternating sign~\cite{the6,HLsim}. Thus,  we assume a vanishing average charge density and take $\sigma_\m=0$. Figure~\ref{fig4}(a) illustrates the polymer translocation rates $R_c$ versus the salt concentration in the reservoir for various polymer charge density values.  At low ion densities, $R_\ce$  is vanishingly small. Above a critical density $\rho_\rb^*$, $R_c$ grows sharply and converges towards the drift velocity $v_{\rm dr}$. One also sees that the capture of polymers with stronger charge occurs at higher salt concentrations, i.e. $|\tau|\uparrow\rho_\rb^*\uparrow$.

In order to understand the physical mechanism behind these features, we plot in Figs.~\ref{fig4}(b) and (c) the polymer potential and velocity profiles. At the salt concentration $\rho_b=0.04$ M where the pore rejects the polymer (purple symbol in Fig.~\ref{fig4}(a)), the electrostatic barrier experienced by the molecule reaches the considerably high value of $V_\p(z_\p)/L_\p\approx9$ $k_{\rm B}T/$nm. To shed light on the origin of this barrier, we note that in a neutral pore where $\phi_\m(r)=0$, the electrostatic interaction energy reduces to the polymer self-energy that can be expressed as the following Fourier integral~\cite{Buy2018I},
\be
\label{31}
\beta\Delta\Omega_\p(l_\p)=l_\p\ell_{\rm B}\tau^2\int_{-\infty}^\infty\mathrm{d}q\frac{2\sin^2(ql_\p/2)}{\pi l_\p q^2}\Delta(q),
\ee
where we introduced the dielectric jump function
\be\label{32}
\Delta(q)=\frac{p_\rb\mathrm{K}_0\left(|q|d\right)\mathrm{K}_1\left(p_\rb d\right)-\gamma |q|\mathrm{K}_1\left(|q|d\right)\mathrm{K}_0\left(p_\rb d\right)}{p_\rb\mathrm{K}_0\left(|q|d\right)\mathrm{I}_1\left(p_\rb d\right)+\gamma |q|\mathrm{K}_1\left(|q|d\right)\mathrm{I}_0\left(p_\rb d\right)},
\ee
with the screening parameter $p_\rb=\sqrt{\kappa^2+q^2}$, the dielectric contrast factor $\gamma=\e_\m/\e_\w$, and the modified Bessel functions $K_{\rm n}(x)$~\cite{math}. The grand potential of Eq.~(\ref{31}) giving rise to the electrostatic barrier $V_p(z_p)$ corresponds to the interaction energy of the polymer with its electrostatic image. At the dilute salt concentration $\rho_\rb=0.04$ M where the highly repulsive image-charge potential $V_\p(z_\p)$ dominates the drift term of Eq.~(\ref{polp}),  the polymer potential $U_\p(z_\p)$ exhibits a minimum followed by an uphill trend and a barrier at $z_\p=L_\m$ (dashed purple curve in Fig.~\ref{fig4}(b)). Due to this barrier, during the polymer capture regime $z_\p<L_\m$, the polymer velocity $v_\p(z_\p)=-\beta D U'_\p(z_\p)$ drops and switches from positive to negative (purple curve in Fig.~\ref{fig4}(c)). The change of the velocity sign indicates polymer trapping by the image-charge barrier at the pore entrance. The system is located in the \textit{barrier-driven} regime.

In strong salt conditions $\kappa l_\p\gg1$ and $\kappa d\gg1$, the polymer grand potential of Eq.~(\ref{31}) can be approximated by
\be\label{33}
\beta\Delta\Omega_\p(l_\p)\approx l_\p\ell_{\rm B}\tau^2\frac{\mathrm{K}_1\left(\kappa d\right)}{\mathrm{I}_1\left(\kappa d\right)}\approx\pi\ell_{\rm B}l_\p\tau^2e^{-2\kappa d}.
\ee
In Fig.~\ref{fig4}(b), one notes that due to the exponential screening of the image-charge barrier formula~(\ref{33}), the increment of the salt concentration from $\rho_\rb=0.04$ M to $0.1$ M reduces the electrostatic potential  from $V_\p(z_\p)/L_\p\approx9$ $k_{\rm B}T/$nm to $\approx3$ $k_{\rm B}T/$nm.  As a result, the drift force on the polymer takes over the electrostatic barrier and the polymer potential $U_\p(z_\p)$ turns to downhill. The polymer is now in the \textit{drift-driven} regime characterized by Eq.~(\ref{34}). Figure~\ref{fig4}(c) shows that this leads to a purely positive velocity, indicating the successful polymer capture and translocation (see also the orange symbol in Fig.~\ref{fig4}(a)).

\begin{figure}
\begin{centering}
\includegraphics[width=.5\linewidth]{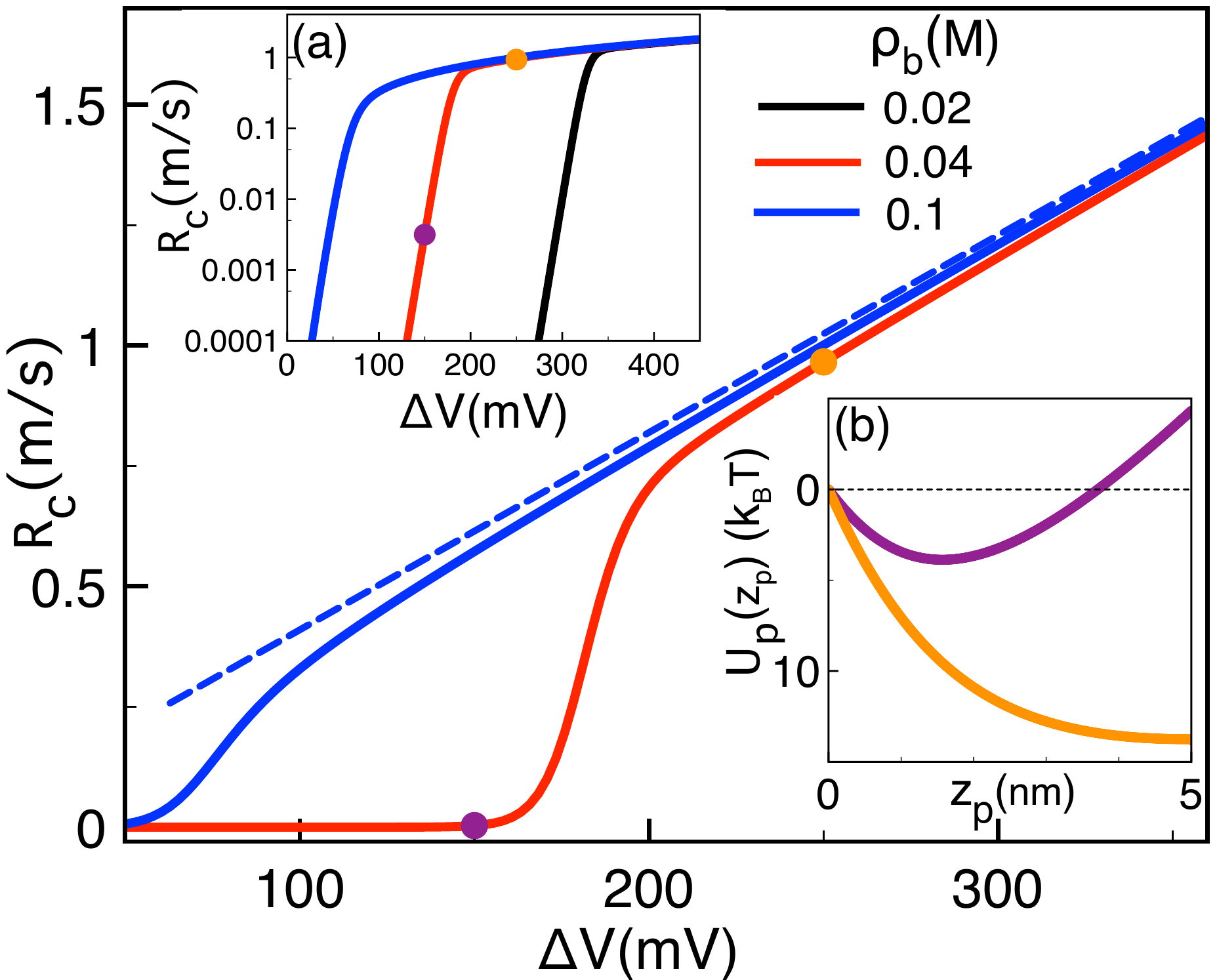}
\caption{(Color online) Main plot: Polymer capture rate $R_\ce$ (solid curves) and drift velocity $v_{\rm dr}$ (dashed curve) against external voltage $\Delta V$ at the salt concentration values indicated in the legend. Insets: (a) the curves of the main plot shown on a logarithmic scale and (b) the polymer potential $U_\p(z_\p)$ at $\Delta V=150$ mV (purple) and $250$ mV (orange). The other parameters are the same as in Fig.~\ref{fig4}.}
\label{fig5}
\end{centering}
\end{figure}

These results indicate that the sharp rise of the polymer capture rates by salt addition originates from the competition between the image-charge barrier and the drift force.  The same competition can indeed allow to understand the turnover in the voltage dependence of experimental polymer translocation rates in $\alpha$HL channels~\cite{e2,e3,e8}.  Figure~\ref{fig5} illustrates this peculiarity at various salt concentration values. In  agreement with the experimental curves of Ref.~\cite{mel}, the translocation rates rise exponentially at low voltages but grow with a weaker slope beyond a crossover voltage $\Delta V^*$. One also notes that added salt reduces this critical voltage, i.e. $\rho_\rb\uparrow\Delta V^*\downarrow$.

The transition in the voltage dependence of the capture rates can be explained in terms of the potential profile $U_\p(z_\p)$ displayed in the inset of Fig. \ref{fig5}. At the voltage $\Delta V=150$ mV (purple curve) located in the exponentially rising regime of the $R_\ce-\Delta V$ curves (purple symbols) the image-charge barrier results in a potential trap to be escaped by thermal fluctuations. At the higher voltage $\Delta V=250$ mV located in the regime where $R_\ce$ increases linearly and gets close to the drift velocity (orange dots), the enhanced drift force takes over the repulsive image-charge barrier and the potential $U_\p(z_\p)$ turns to downhill (orange curve in Fig. \ref{fig5}). This indicates that the non-uniform voltage dependence of the translocation rates is a consequence of the transition from the barrier to drift-dominated polymer transport regime. 

\subsection{Limitation of the stiff polymer approximation}

The main approximation of the electrohydrodynamic translocation theory presented herein is the modeling of the polymer as a rigid rod. We have shown that despite this approximation, accurate and detailed modeling of the pore electrohydrodynamics enables the theory to quantitatively explain several results obtained in many translocation experiments. This said, because the entropic cost of polymer translocation originating from conformational polymer fluctuations becomes important beyond the DNA persistence length $L_\p\gtrsim50$ nm, the rigid rod approximation limits the quantitative predictive power of the electrohydrodynamic theory to polymer sequences whose size is comparable with the length of solid-state pores. Thus, if one wishes to consider the translocation of long polymer sequences, the inclusion of conformational polymer fluctuations becomes unavoidable. We are currently working in this direction.

The theoretical or even numerical consideration of electrohydrodynamic forces on a fluctuating polymer presents itself as an almost untractable task. However, the rigid rod approximation can be relaxed in the opposite regime of polymers much longer than the nanopore. In this configuration where the force induced by the pore electrohydrodynamics on the polymer can be considered to be local, one can absorb the electrohydrodynamic forces on the DNA into an effective external force $f$ exerted solely on the polymer portion located in the pore, and the effective pore friction $\eta_{\rm p}$. This simplification allows to bypass the detailed description of the pore electrohydrodynamics, thereby enabling the accurate consideration of the polymer conformations bringing a major contribution to the translocation of long polymers. The next section of our article is devoted to this type of configurational translocation approach called the \textit{iso-flux tension propagation theory}.

\section{Iso-flux tension propagation (IFTP) theory for the translocation of long polymers} 
\label{IFTP}

This section is devoted to the tension propagation theory of polymer translocation through a nanopore.
First, the theoretical model is introduced. Then to show the validity of 
the tension propagation theory the dynamics of the polymer translocation process is examined at the monomer level 
by looking at the waiting time distribution, that is the time each bead spends at the pore during the course 
of translocation. In the next subsection the scaling form of the translocation time $\tau_2$ (cf. Eq. (29)), which is the time that the 
chain needs to completely pass through the nanopore, is obtained for both pore-driven and end-pulled cases.
We note that the theory presented here does not include any specific capture or escape processes, but assumes that the
translocation starts with the pore being already filled and and stops when the {\it cis} side has no monomers left. This
corresponds to the assumption that the pore thickness $L_{\rm m} \ll L_{\rm p}$.
Finally, in the last subsection we discuss the application of the theory to semi-flexible and rodlike polymers.

\subsection{Coarse-grained polymer model} \label{model}

Following our previous works, in this section we denote the polymer contour length by $N_0$, and the
translocation time $\tau = \tau_2$, since in the theory here $\tau_1 = \tau_3 = 0$ corresponding to the thin pore approximation
without any specific capture or trapping processes.
For brevity, dimensionless units denoted by tilde are used 
as $\tilde{Z} \equiv Z / Z_{\textrm{u}}$, with the units of length $s_{\textrm{u}} \equiv {\textrm{a}}$, 
time $t_{\textrm{u}} \equiv \eta a^2 / (k_{\textrm{B}} T)$, 
force $f_{\textrm{u}} \equiv k_{\textrm{B}} T/a$, velocity $v_{\textrm{u}} \equiv a/t_{\textrm{u}} = k_{\textrm{B}} T/(\eta a)$,  
friction $\Gamma_{\textrm{u}} \equiv \eta$, and monomer flux $\phi_{\textrm{u}} \equiv k_{\textrm{B}} T/(\eta a^2)$.
Here $a$ is the segment length, $T$ is the temperature of the system, $k_{\textrm{B}}$ is the Boltzmann constant, 
and $\eta$ is the friction of the solvent per monomer. Variables without tilde are expressed in Lennard-Jones units 
(for details see Refs.~\cite{jalal2017SR,jalal2017EPL}).

\begin{figure} 
\begin{centering}
\includegraphics[width=1.\textwidth]{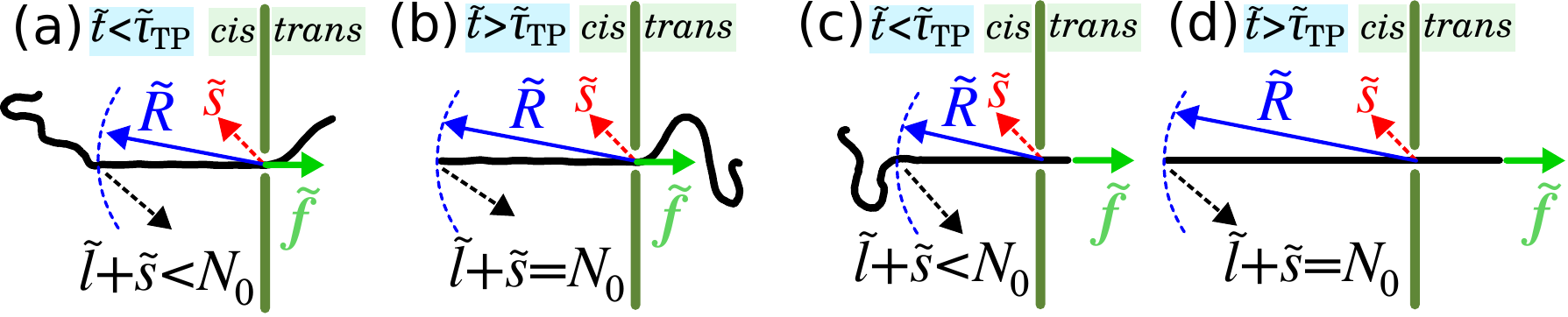}
\caption{(a) Schematic of the pore-driven translocation process during tension propagation (TP) stage, i.e. 
$\tilde{t} < \tilde{\tau}_{\rm{TP}}$, for the SS regime.
The external driving force $\tilde{f}$ acts only on the monomer(s) at the pore towards the {\it trans} side. 
$N_0$ is the contour length of polymer, and $\tilde{s}$ is the number of segments that have 
already been translocated into the {\it trans} side. $\tilde{l} + \tilde{s}$ is the number of 
beads influenced by the tension force which is less than $N_0$ during the TP stage.
$\tilde{R}$ denotes the location of the tension front.
(b) Translocation process for the SS regime in the post propagation (PP) stage when the tension has reached the
chain end and after that, i.e. $\tilde{l} + \tilde{s} = N_0$. (c) The same as (a) but for the end-pulled polymer 
translocation process where the external driving force acts only on the head monomer
in the direction perpendicular to the membrane from {\it cis} to {\it trans} side. 
(d) The same as (b) but for the end-pulled case.} 
\label{fig6}
\end{centering}
\end{figure}

During the process of polymer translocation the driving force may either act on the monomer(s) 
inside the pore (pore-driven case) or on the head monomer of the polymer (end-pulled case). For both pore-driven and 
end-pulled cases when the driving force is switched on a tension front starts to propagate along the backbone of 
the chain. Consequently, the {\it cis} part of the chain can be divided into two parts, 
mobile and immobile ones (see Figs.~\ref{fig6}(a) and (c)) \cite{sakaue2007}. 
Indeed, the part of the chain which experiences tension 
is mobile and has non-zero net velocity and the rest of the chain is in an immobile equilibrium state 
with zero average velocity. In the pore-driven case the velocity of the mobile part is towards the pore 
(see Fig.~\ref{fig6}(a)) while for the end-pulled case it is in the direction of 
the driving force (see Fig.~\ref{fig6}(c)). The boundary between the mobile and immobile parts is called the tension 
front that is located on the {\it cis} side for both pore-driven and end-pulled cases (see Fig.~\ref{fig6}).
Both processes comprise two stages, 
the tension propagation (TP) and post propagation (PP) ones. During the TP stage the tension has not reached 
the chain end (see Figs.~\ref{fig6}(a) and (c)) while in the PP stage the whole chain has already been influenced 
by the tension (see Figs.~\ref{fig6}(b) and (d)). 

The shape of the mobile subchain depends on the strength of the driving force. For the pore-driven case
in the limit of weak ($ N_0^{-\nu} \ll \tilde{f} \ll 1$) and moderate forces ($1 \ll \tilde{f} \ll   N_0^{\nu}$) 
the mobile subchain is reminiscent of the trumpet (TR) and stem-flower (SF) configurations, respectively, while in the limit of 
very strong force ($N_0^{\nu} \ll \tilde{f} $), in the strong stretching (SS) regime, the mobile subchain is fully straightened 
\cite{rowghanian2011,ikonen2012a,ikonen2012b,jalal2014}. 
Here, $N_0$ is the contour length of the polymer, $\nu$ is the Flory exponent which is $3/4$ and 0.588 for 
excluded volume chains in a good solvent is 2D and 3D, respectively, and $\tilde{f}$ is the 
external driving force that acts on the monomers inside the pore for the pore-driven case or on the head monomer 
for the end-pulled case as depicted in Fig. \ref{fig6}.
For the end-pulled case as the mobile part extends to the {\it cis} and the {\it trans} sides, 
the dynamics of the chain is more complicated than that of the pore-driven case.
For the {\it cis} side mobile subchain the same scenario as for the pore-driven case is valid here, but instead of the 
driving force, $\tilde{f}$, one needs to measure the value of the mediated tension force at the pore, $\tilde{f}_{\rm{p}}$.
Thus, the shape of the mobile subchain in the {\it cis} side fits into the TR and SF regimes if 
$(N_0-\tilde{s})^{-\nu} \ll \tilde{f}_{\rm{p}} \ll 1$ and $1 \ll \tilde{f}_{\rm{p}} \ll (N_0-\tilde{s})^{\nu}$, respectively, 
where $\tilde{s}$ is the translocation coordinate that is the length of the subchain in the {\it trans} side. 
The {\it cis} side mobile subchain is fully straightened if $ (N_0-\tilde{s})^{\nu} \ll \tilde{f}_{\rm{p}} $.
In the other hand, for the end-pulled case the {\it trans} side mobile subchain shape can be either TR, 
SF or fully straightened according to the strength of the driving force. For example the mobile 
subchain in the {\it trans} side is fully straightened if $ N_0 \ll \tilde{f} $ \cite{jalal2017EPL}.

To study polymer translocation through a nanopore, similar to Refs.~\cite{ikonen2012a,ikonen2012b,jalal2014} 
the basic framework of Brownian dynamics (BD) in the overdamped limit is employed. According to BD the equation 
of motion for the translocation coordinate $\tilde{s}$ is written as
\begin{equation}
 \tilde{\Gamma} (\tilde{t}) \frac{d \tilde{s}}{ d \tilde{t}} =
(1- \gamma ') \bigg[ \frac{1}{N_0 - \tilde{s}} - \frac{1}{\tilde{s}} \bigg] 
+ \tilde{f} + \tilde{\zeta} (\tilde{t}) \equiv  \tilde{f}_{\textrm{tot}},
\label{BD_equation}
\end{equation}
where $\tilde{\Gamma} (\tilde{t})$ is the total effective friction, $\gamma '$ is the surface exponent 
which is $\gamma ' \approx ~ 0.95$ and $\approx~ 0.69$ for self-avoiding chains in 2D and 3D, 
respectively, and $\gamma '=0.5$ for ideal chains, 
$\tilde{\zeta} (\tilde{t})$ is Gaussian white noise that satisfies $\langle \zeta (t) \rangle = 0$ and 
$\langle \zeta (t) \zeta (t') \rangle = 2 \Gamma (t) k_{\textrm{B}} T \delta (t - t ')$,
and $\tilde{f}_{\textrm{tot}}$ is the total force.
The effective friction can be written as a sum of the friction due to the mobile part of the chain 
and the pore frictions. For the pore-driven case the effective friction is 
$\tilde{\Gamma} (\tilde{t}) = \tilde{\eta}_{\textrm{cis}} (\tilde{t}) + \tilde{\eta}_{\rm{p}}$, where 
$\tilde{\eta}_{\textrm{cis}} (\tilde{t})$ is the friction due to the movement of mobile subchain in the solvent. 
For the end-pulled case the effective friction is written as 
$\tilde{\Gamma} (\tilde{t}) = \tilde{\eta}_{\textrm{cis}} (\tilde{t}) + \tilde{\eta}_{\textrm{TS}} (\tilde{t}) + \tilde{\eta}_{\rm{p}}$,
where $\tilde{\eta}_{\textrm{cis}} (\tilde{t}) $ and $ \tilde{\eta}_{\textrm{TS}} (\tilde{t})$ are frictions due to the movement of 
the mobile parts of the chain in the {\it cis} and in the {\it trans} sides inside the solvent, respectively. 
The index TS is an abbreviation for the {\it trans} side.
It should be mentioned that for the pore-driven case of a flexible chain the dynamical {\it trans} side friction can 
be adsorbed into the pore friction as it just contributes a constant factor to it \cite{ikonen2012a,ikonen2012b,ikonen2013}.
We also note that the term proportional to $1-\gamma'$ arises from the equilibrium entropy of the 
chain (for a fixed $\tilde{s}$) and is
small enough to be neglected in the SS regime. It should be noted that it is not fully consistent with the propagation
of the tension front on the {\it cis} side even for the pore-driven case.

Equation \eqref{BD_equation} gives the time evolution of the translocation coordinate, $\tilde{s}$, provided that the 
effective friction, $\tilde{\Gamma} (\tilde{t})$, is known. Indeed, the physics of tension propagation theory is embedded in 
$\tilde{\Gamma} (\tilde{t})$. To find the effective friction that is the combination of the mobile subchain and pore frictions, 
one needs to find the time evolution of the tension front, which gives the dynamics of the friction due to the mobile subchain. 
To this end, similar to Ref.~\cite{rowghanian2011}, we assume that the flux of monomers in the mobile domain, 
$\tilde{\phi} = d \tilde{s} / d \tilde{t}$, is constant in space but evolves in time (the iso-flux assumption). 
The tension front is located at the distance $\tilde{x} = - \tilde{R}$ from the pore on the {\it cis} side.
The tension force at the distance $\tilde{x}$ from the pore is obtained by integrating the force-balance equation
for a differential element $d \tilde{x}$ located between $\tilde{x}$ and $\tilde{x} + d \tilde{x}$, as
$\tilde{f} ( \tilde{x} , \tilde{t} ) = \tilde{f}_0 - \tilde{\phi} (\tilde{t}) \tilde{x} $. For the pore-driven case
the integration is performed from pore to $\tilde{x}$ and 
$\tilde{f}_0 \equiv \tilde{f}_{\textrm{tot}} - \tilde{\eta}_{\textrm{p}} \tilde{\phi} (\tilde{t}) $, while
for the end-pulled case the integration is from head monomer to pore and then from pore to $\tilde{x}$
and 
$\tilde{f}_0 \equiv \tilde{f}_{\textrm{tot}} - \tilde{\eta}_{\textrm{p}} \tilde{\phi} (\tilde{t})
- \tilde{\eta}_{\textrm{TS}} \tilde{\phi} (\tilde{t}) $. Here, in the SS regime $\tilde{\eta}_{\textrm{TS}} = \tilde{s}$ for the end-pulled case. 
Indeed, by integration of the force balance equation over the whole mobile domain together with the definition of 
the tension front, where the tension force vanishes, the equation of motion for the monomer flux is written as
\begin{eqnarray}
\tilde{\phi} (\tilde{t}) &=& \frac{\tilde{f}_{\textrm{tot}} (\tilde{t})}
{ \tilde{R} (\tilde{t}) + \tilde{\eta}_{\textrm{p}} }, \hspace{1.45cm} \textrm{pore-driven}; \nonumber\\
\tilde{\phi} (\tilde{t}) &=& \frac{\tilde{f}_{\textrm{tot}} (\tilde{t})}
{ \tilde{R} (\tilde{t}) + \tilde{\eta}_{\textrm{p}} + \tilde{\eta}_{\textrm{TS}} }, \hspace{0.5cm} \textrm{end-pulled}.
\label{phi_equation}
\end{eqnarray}
Then the effective friction is expressed by using Eqs. \eqref{BD_equation}, \eqref{phi_equation} and 
the definition of the monomer flux, $\tilde{\phi}\equiv d\tilde{s}/d\tilde{t}$, as
\begin{eqnarray}
\tilde{\Gamma} (\tilde{t}) &=& \tilde{R}(\tilde{t}) + \tilde{\eta}_{\textrm{p}}, \hspace{1.99cm} \textrm{pore-driven}; \nonumber\\
\tilde{\Gamma} (\tilde{t}) &=& \tilde{R}(\tilde{t}) + \tilde{\eta}_{\textrm{p}} + \tilde{\eta}_{\textrm{TS}},
\hspace{1.05cm} \textrm{end-pulled}.
\label{Gamma_equation}
\end{eqnarray}
Time evolution of $\tilde{s}$ is given by Eqs. \eqref{BD_equation}, \eqref{phi_equation} and
\eqref{Gamma_equation}, but to have a full solution one still needs the equation of motion for
the location of the tension front, $\tilde{R}$, which is obtained in the TP and PP stages separately. 
In the TP stage where the tension front has not reached the chain end, for flexible chain one can write $\tilde{R} = A_{\nu} N^{\nu}$,
where $N= \tilde{l} + \tilde{s}$ is the number of segments that already influenced by the tension (see Fig.~\ref{fig6}(a)) 
and $\tilde{l}$ is the number of segments in the mobile domain in the {\it cis} side. Here we only present the time evolution of the 
tension front for the strong stretching (SS) regime, where the force is very strong, 
and $\tilde{l} = \tilde{R}$.
To study the other TR and SF regimes a similar procedure is employed \cite{jalal2014}.
Inserting $\tilde{N}$ inside the equation above for $\tilde{R}$ and performing time derivation,
the equation of motion for the tension front for TP stage is obtained as
\begin{equation}
\dot{\tilde{R}} (\tilde{t}) = \frac{ \nu A_{\nu}^{1/\nu}  \tilde{R} (\tilde{t})^{\frac{\nu -1}{\nu}} \tilde{\phi} (\tilde{t})  }
{ 1- \nu A_{\nu}^{1/\nu}  \tilde{R} (\tilde{t})^{\frac{\nu -1}{\nu}} }.
\label{evolution_of_R_TP}
\end{equation}
In the PP stage the tension force has already reached the chain and therefore $N = \tilde{l} + \tilde{s} = N_0 $. 
By substituting $\tilde{l} = \tilde{R} $ (in the SS regime) in the above relation and taking the time derivative,
the equation for the time evolution of the tension front is written as
\begin{equation}
\dot{\tilde{R}} (\tilde{t}) = \tilde{\phi} (\tilde{t}) .
\label{evolution_of_R_PP}
\end{equation}
The time evolution of the tension front for the end-pulled case in the SS regime for TP and PP stages is the same as of 
pore-driven case, i.e. Eqs. \eqref{evolution_of_R_TP} and \eqref{evolution_of_R_PP}.

To find a full solution of the iso-flux tension propagation (IFTP) model for the TP stage Eqs. (\ref{BD_equation})-(\ref{evolution_of_R_TP})
must be consistently solved, while for the PP stage 
one needs to solve Eqs.~\eqref{BD_equation}-\eqref{Gamma_equation}, and \eqref{evolution_of_R_PP}.
It should be mentioned that to improve the quantitative accuracy of the IFTP theory the distribution of the initial configurations of 
the chain can be incorporated into the model through $A_{\nu}$ in $\tilde{R} = A_{\nu} N^{\nu}$. The details
are in Ref. \cite{jalal2014}.
Moreover, modified versions of IFTP theory have been employed to study the translocation of a semi-flexible or stiff polymer 
through a nanopore \cite{jalal2017SR} (to be discussed later) as well as polymer translocation through a flickering nanopore under an alternating 
driving force \cite{jalal2015}.

\subsection{Waiting time distribution} \label{waiting_time}

\begin{figure}
\begin{centering}
\includegraphics[width=0.4\textwidth]{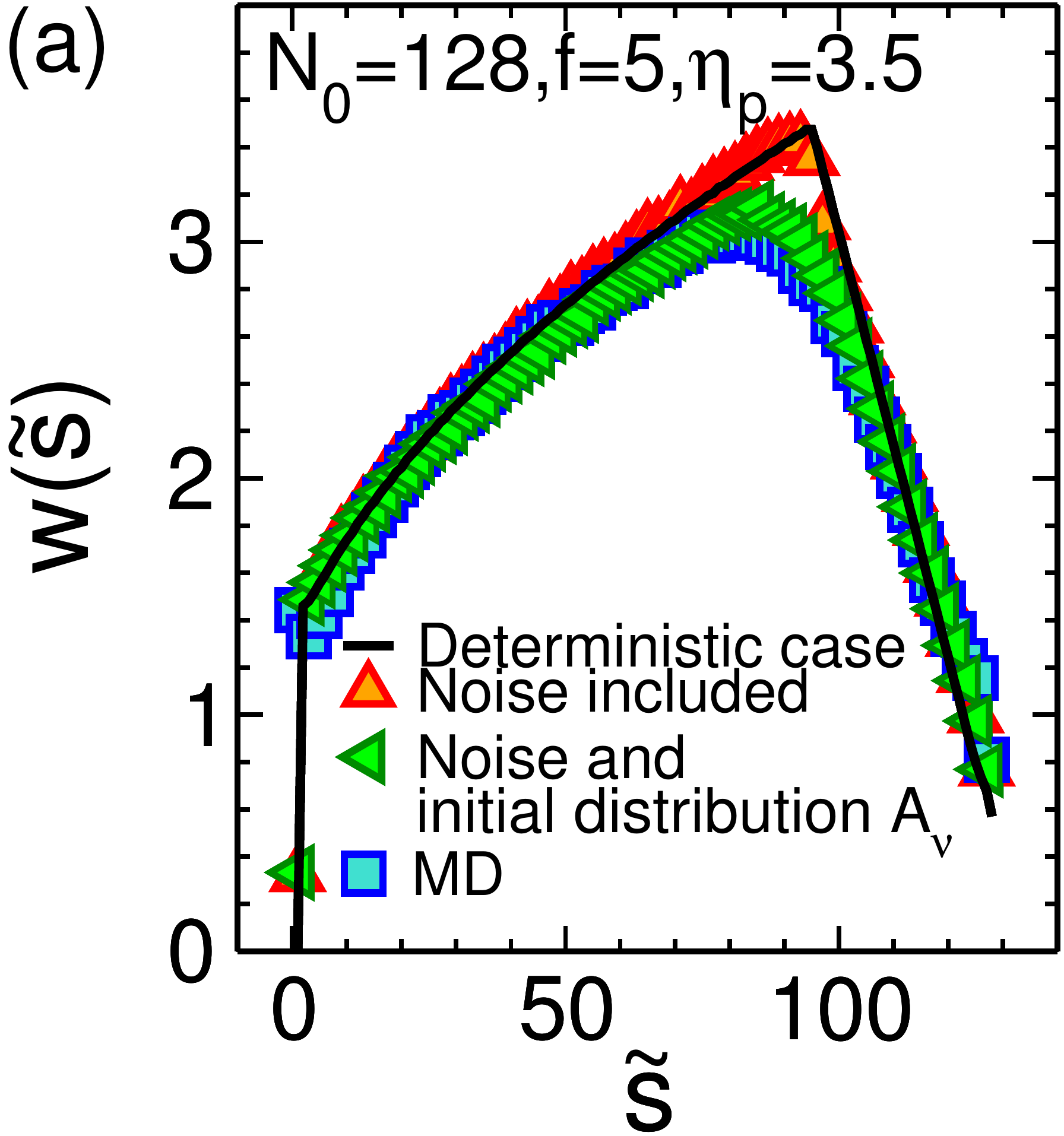}
\includegraphics[width=0.4\textwidth]{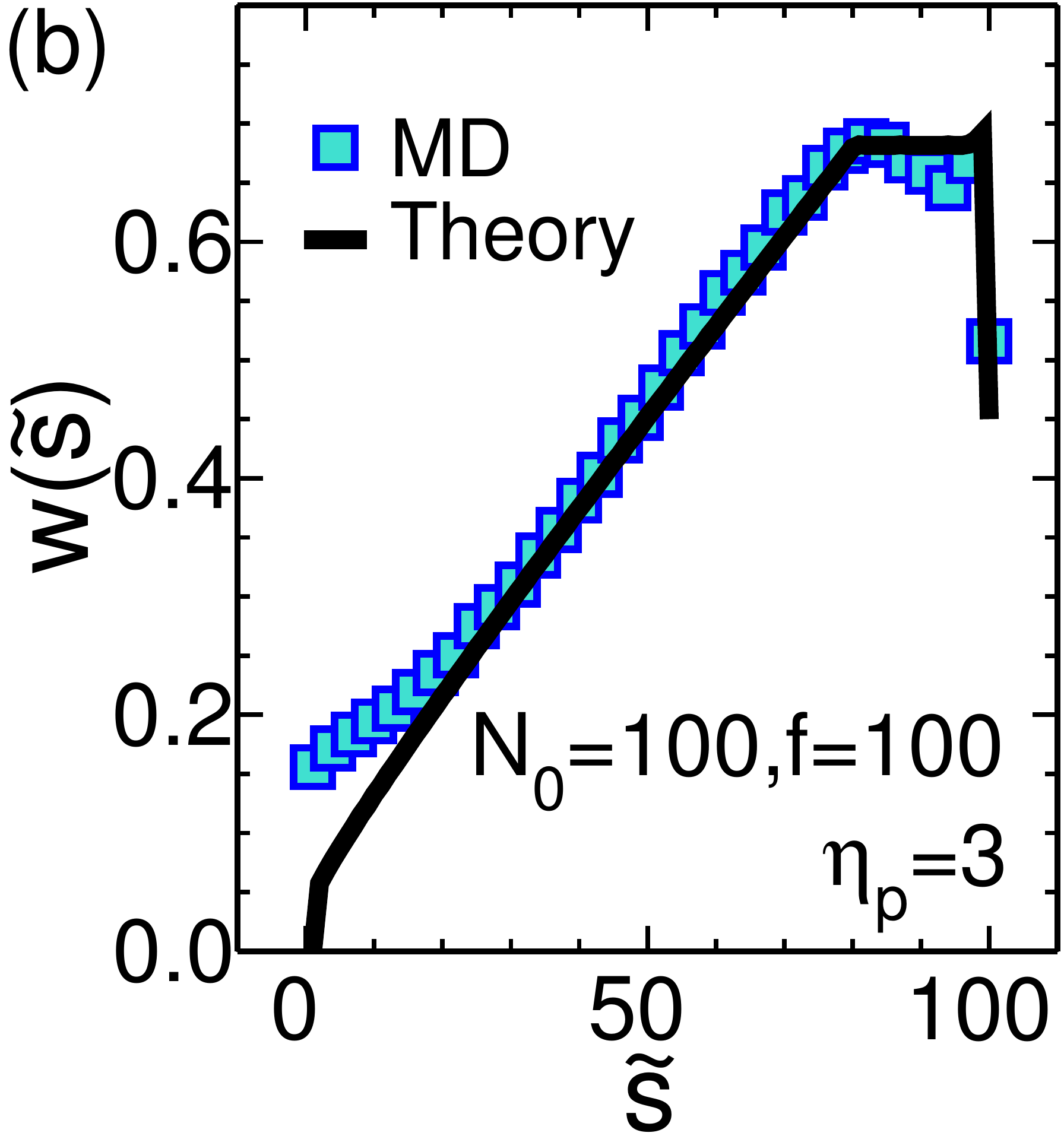}
\caption{(a) The waiting time (WT) $w (\tilde{s})$ as a function of $\tilde{s}$ for the pore-driven case. 
The chain length is $N_0 =128$, the external driving force at the pore is $f = 5$, and the pore friction 
in the IFTP theory is ${\eta}_{\rm{p}} = 3.5$. The black curve presents the deterministic case with fixed $A_{\nu}= 1.15$, 
where neither the thermal fluctuations nor the distribution of the initial configurations 
of the chain have been taken into account. The red up triangles show the WT when the force is sampled
from the proper distribution but $A_{\nu}= 1.15$ is deterministic. Green left triangles are the WT data when 
both the force and $A_{\nu}$ have been sampled from their distributions, and finally MD simulation data are blue squares.
(b) The same as (a) but for the end-pulled case. Black curve is for deterministic case while the blue 
squares represent the MD simulation data. Here, the chain length is $N_0 = 100$, the external driving force
is $f = 100$, and the pore friction in the IFTP theory is ${\eta}_{\rm{p}} = 3$.}
\label{WT_figure}
\end{centering}
\end{figure}

The waiting time (WT), which is the time that each bead spends at the pore during the course of translocation,
is an important quantity that can reveal the dynamics of the process at the monomer level.
Figure \ref{WT_figure}(a) shows the WT as a function of $\tilde{s}$, the translocation coordinate, 
for the pore-driven polymer translocation. The chain length is $N_0 =128$,
the external driving force at the pore is $f = 5$, and the pore friction in the IFTP theory is 
$\eta_{\rm{p}} = 3.5$. The black curve presents the deterministic case. The red up triangles show the WT when force is chosen 
randomly but $A_{\nu}= 1.15$ is deterministic. Green left triangles are devoted to the WT when 
both the force and $A_{\nu}$ are stochastic, and finally MD simulation data are shown in blue squares.
The stochastic sampling of the initial configurations of the chain smoothens the transition from the TP to the PP stage. 
In MD simulations (blue squares) the same feature is seen where the initial configurations are sampled by thermalizing the chain
before each simulation trajectory. It is clear there is a very good quantitative 
agreement between the stochastically augmented IFTP theory and MD simulations. 
Panel (b) is the same as panel (a) but for the end-pulled case. Black curve is for deterministic case while the blue 
squares show the MD simulation data. Here, the chain length is $N_0 = 100$, the external driving force
is $f = 100$, and the pore friction in the IFTP theory is ${\eta}_{\rm{p}} = 3$.

For both pore-driven and end-pulled cases, it can be seen in Fig. \ref{WT_figure} that the translocation process is a far-from-equilibrium
process in the sense that the conformations of the chain do not correspond to linear response or quasi-equilibrium ones. For pore-driven case, Fig. \ref{WT_figure}(a), in the TP stage, where the tension is still propagating along the backbone 
of the chain, the number of mobile monomers in the {\it cis}
side is increasing. Therefore the friction is growing and consequently WT increases monotonically until 
it gets its maximum when the tension reaches the end of the chain.
Then in the second PP stage as the time passes the number of mobile monomers in the {\it cis} side decreases,
which means the friction due to the mobile part of the chain decreases too, and WT decreases.
For the end-pulled case, Fig. \ref{WT_figure}(b), in the TP stage similar to the pore-driven case WT increases.
However, in the PP stage WT is almost constant. This is because in the SS regime both subchains in the {\it cis} and in the {\it trans} 
sides are fully straightened (rodlike) therefore the friction due to the movement of the whole chain in the solvent remains
approximately constant.

\subsection{Scaling of the translocation time for a flexible polymer} \label{scaling_flexible_chain}

A fundamental quantity characterizing the polymer translocation process is the translocation time,
which is the time that a chain needs to pass through the nanopore. Here, we consider only the case where
 $\tau = \tau_2$ and show how 
the scaling form of the translocation time can be extracted analytically from the IFTP theory in the SS regime. 
The same approach can be applied for the TR and SF regimes \cite{jalal2014,jalal2015,jalal2017SR,jalal2017EPL,jalal2018JPC}.

To obtain an analytical form of the translocation time in the SS regime we use an approximation and we only take into account 
the contribution of the external driving force to the equation of motion for the translocation coordinate, 
i.e. $\tilde{f}_{\rm{tot}} \approx \tilde{f}$. Then for the pore-driven 
case, combining $\tilde{\phi} = \tilde{f} /\big( \tilde{\eta}_{\rm{p}} + \tilde{R} \big) $ with the definition of the 
monomer flux, $\tilde{\phi} = d \tilde{s} / d \tilde{t} $, together with the mass conservation in the TP stage, 
$N= \tilde{l} + \tilde{s}$, by integration of $N$ from 0 to $N_0$ the TP time reads as
$\tilde{\tau}_\mathrm{TP} =  \big[ \int_0^{N_0} \tilde{R}(N) dN + \tilde{\eta}_{\rm{p}} N_0 \big] /\tilde{f}
- \Delta \tilde{\tau}$,
where $\Delta \tilde{\tau} = \big[ \tilde{\eta}_{\rm{p}} \tilde{R} (N_0) + \tilde{R}^2 (N_0) /2  \big] / \tilde{f}$.
The PP time is obtained by integrating $\tilde{R}$ from $\tilde{R} (N_0)$ to 0 as $\tilde{\tau}_\mathrm{PP} = \Delta \tilde{\tau}$.
At the end, the whole translocation time, $ \tilde{\tau} = \tilde{\tau}_\mathrm{PP} + \tilde{\tau}_\mathrm{TP}$, 
is written as $\tilde{\tau} = \big[ \int_0^{N_0} \tilde{R}(N) dN + \tilde{\eta}_{\rm{p}} N_0 \big] / \tilde{f}$ 
with the scaling form
\begin{equation}
\tilde{\tau} = 
\frac{1}{\tilde{f}} \bigg[ \frac{ A_{\nu} N_0^{1+\nu} }{1+\nu} + \tilde{\eta}_{\rm{p}} N_0 \bigg].
\label{whole_time_Pore_driven}
\end{equation}
To obtain the scaling of the translocation time for the end-pulled case the same procedure 
is applied, but now the monomer flux is 
$\tilde{\phi} = \tilde{f} /\big( \tilde{\eta}_{\rm{p}} + \tilde{R} + \tilde{\eta}_{\rm{TS}} \big) $, 
where $\tilde{\eta}_{\rm{TS}}$ is the {\it trans} side friction and for the SS regime is $\tilde{\eta}_{\rm{TS}} = \tilde{s}$.
The whole translocation time for the end-pulled case is given by 
$\tilde{\tau} = \frac{1}{\tilde{f}} \big[ \int_0^{N_0} \tilde{R}(N) dN + \tilde{\eta}_{\rm{p}} N_0 \big] + \tilde{\tau}_\mathrm{TS}$,
where $\tilde{\tau}_\mathrm{TS} = N_0^2 / (2\tilde{f})$ is the contribution of friction due to the fully straightened {\it trans}
side subchain. Thus
\begin{equation}
\tilde{\tau} = \frac{1}{\tilde{f}} \bigg[ \frac{ A_{\nu} N_0^{1+\nu} }{1+\nu} + \tilde{\eta}_{\rm{p}} N_0 + \frac{ N_0^2 }{2} \bigg].
\label{whole_time_End_pulled}
\end{equation}

\begin{figure}[t]\begin{center}
    \begin{minipage}[b]{0.4\textwidth}\begin{center}
        \includegraphics[width=0.95\textwidth]{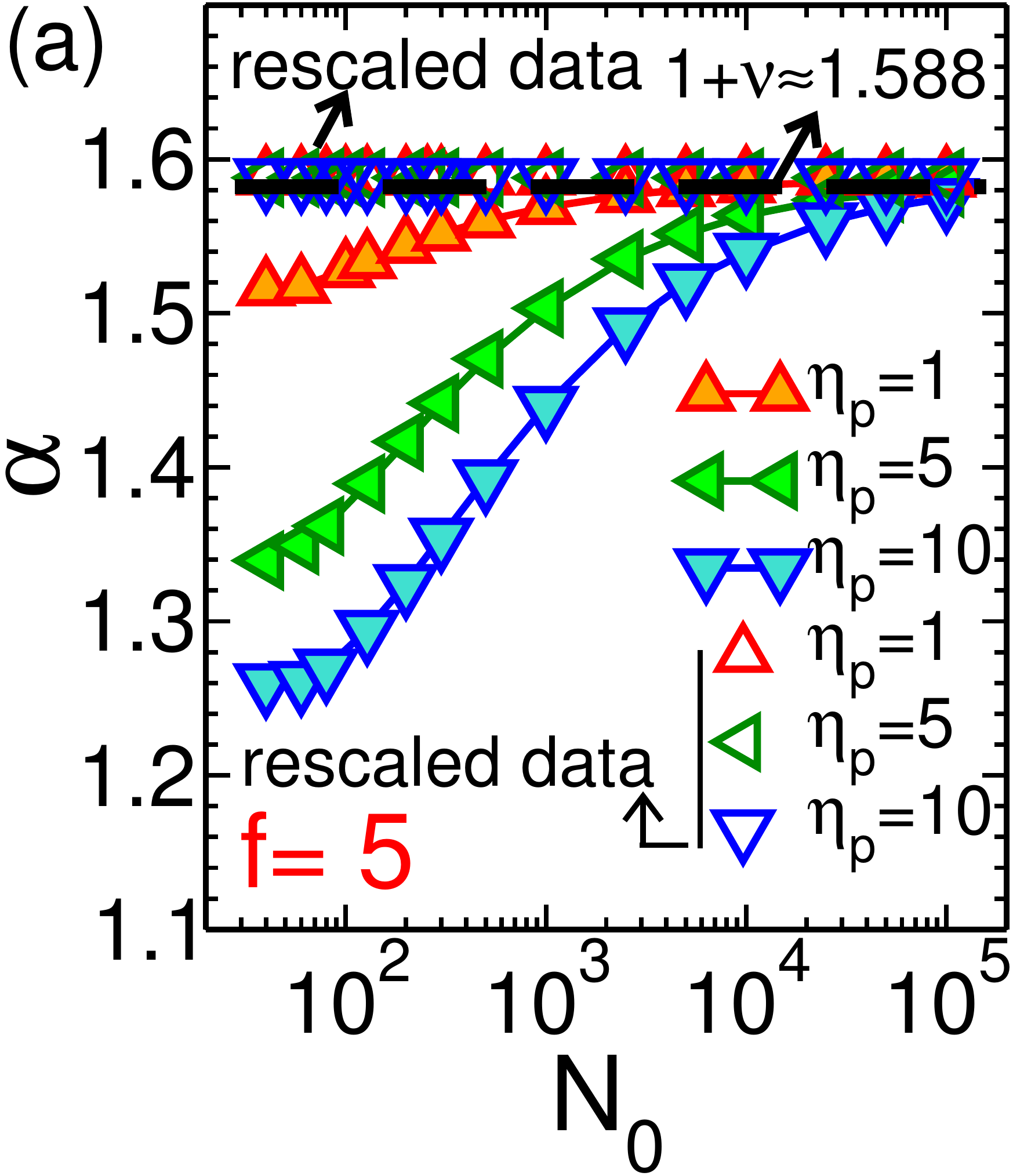}
    \end{center}\end{minipage} \hskip+0.45cm
    \begin{minipage}[b]{0.4\textwidth}\begin{center}
        \includegraphics[width=0.90\textwidth]{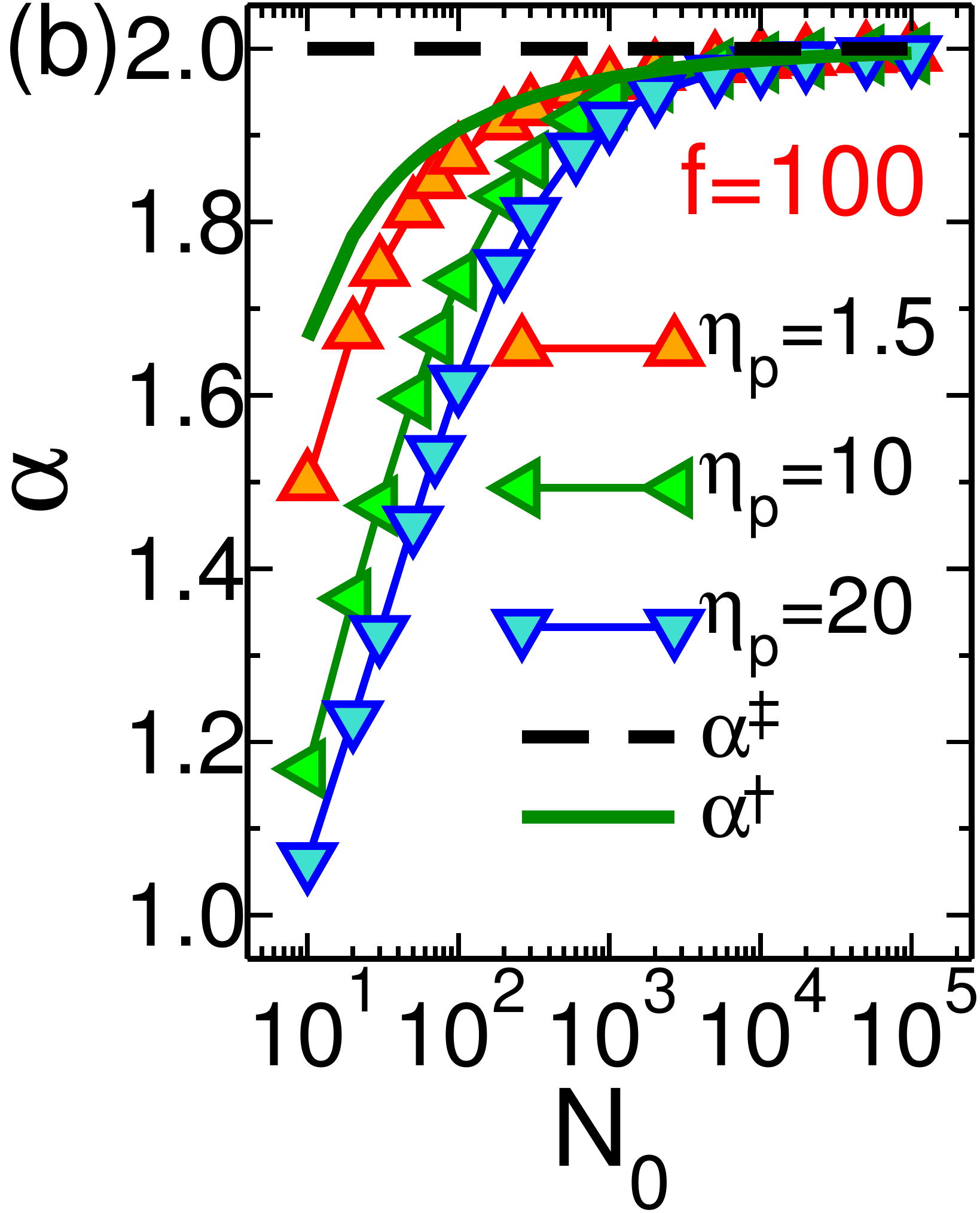}
    \end{center}\end{minipage}
\caption{(a) The translocation exponent $\alpha$ for the pore-driven case based on the IFTP theory
as a function of $N_0$ for various values of the pore friction ${\eta}_{\rm{p}} = 1, 5$ and 10. 
Here the external driving force is $f= 5$, and $A_{\nu}= 1.15$. 
The rescaled exponent curves are presented for different values of $\eta_{\rm{p}}$. They collapse 
on a single curve as denoted by rescaled data in the figure, i.e. $\alpha^{\dag}= 1 + \nu= 1.588$.
(b) Similar to (a) but for end-pulled case and for various values of the pore friction 
${\eta}_{\rm{p}} = 1.5, 10$ and 20. Here the external driving force is $f= 100$. 
The green solid line presents the rescaled translocation exponent $\alpha^{\dag}$ while the black dashed curve shows 
the rescaled exponent $\alpha^{\ddag}$ (for definitions of $\alpha^{\dag}$ and $\alpha^{\ddag}$ see the text).} 
\label{exponent_figure}
\end{center}
\end{figure}

In Fig. \ref{exponent_figure}(a) we plot the translocation exponent $\alpha$, 
which is defined as $\tilde{\tau} \propto N_0^{\alpha}$, for 
the pore-driven polymer translocation based on the deterministic IFTP theory as a function of $N_0$ for various values of the 
pore friction ${\eta}_{\rm{p}} = 1, 5$ and 10. Here the external driving force is $f= 5$, and $A_{\nu}= 1.15$. 
As can be seen in the short chain limit the values of the effective translocation exponents for different pore friction coefficients
are different due to a competition between the two terms in the right hand side of Eq. \eqref{whole_time_Pore_driven}.
In the long chain limit where the first term due to the mobile subchain friction dominates, the values of the translocation 
exponents for different pore frictions are the same, as depicted in the figure by $1 +\nu$.
To show the effect of the pore friction on the translocation exponent, the rescaled exponent curves are presented for 
different values of $\eta_{\rm{p}}$. As can be seen all of them collapse on a single curve as denoted by rescaled data 
in the figure, i.e. $ 1 + \nu = 1.588$. For the pore-driven case the rescaled exponent is defined as 
$\tilde{\tau} - \tilde{\eta_{\rm{p}}} N_0 /\tilde{f} \propto N_0^{\alpha^{\dag}}$
and shows only the {\it cis} side friction contribution to the dynamics of the translocation process.
Figure \ref{exponent_figure}(b) is similar to Fig. \ref{exponent_figure}(a) but for the end-pulled case, and for various values of the pore friction 
${\eta}_{\rm{p}} = 1.5, 10$ and 20. Here the external driving force is $f= 100$ acts on the head monomer of the polymer.
The green solid line presents the rescaled translocation exponent $\alpha^{\dag}$, which is defined as 
$\tilde{\tau} -\tilde{\eta_{\rm{p}}} N_0 /\tilde{f}  
\propto N_0^{\alpha^{\dag}}$. The black dashed curve shows the rescaled exponent $\alpha^{\ddag} = 2$
with the definition of 
$\tilde{\tau} - \big[ \int_0^{N_0} \tilde{R}(N) dN - \tilde{\eta}_{\rm{p}} N_0 \big] / \tilde{f} 
= \tilde{\tau}_\mathrm{TS} \propto N_0^{\alpha^{\ddag}}$. 

\subsection{Scaling of the translocation time for a stiff polymer} \label{scaling_stiff_chain}

In this subsection the scaling form of the translocation time for a stiff chain is briefly discussed  
for both pore-driven and end-pulled polymer translocation processes. The complete theory for semi-flexible
chains can be found in Ref. \cite{jalal2017SR}.
The end-to-end distance of a very stiff chain (rod-like limit) is given by $\tilde{R} (N) = N$. For both pore-driven
and end-pulled translocation processes of the stiff polymer the number of mobile monomers on the {\it cis} side is 
$\tilde{l} = \tilde{R}$, while on the {\it trans} side it is given by $\tilde{s}$. As the chain is very stiff 
the total translocation time is much larger than the TP time, i.e. 
$\tilde{\tau} \gg \tilde{\tau}_{\textrm{tp}}$, and therefore it is a very good approximation to ignore the TP stage. 
During the PP stage as the tension has already reached the chain end 
$N= \tilde{s}+ \tilde{l}=N_0$, and one sets the condition $dN/d\tilde{t} = 0$. The PP time, which is equal to the 
total translocation time, is obtained by integrating $\tilde{R}$ from 
$\tilde{R} (N_0)$ to zero. Then, the total translocation time is written as
$\tilde{\tau} = \tilde{\tau}_\mathrm{pp} = \int_{0}^{\tilde{R}_{N_0}} 
d \tilde{R}  \big[ \tilde{R} + \tilde{\eta}_{\textrm{p}} + \tilde{\eta}_{_\textrm{TS}} (\tilde{t}) ] / \tilde{f}$.
Knowing $\tilde{\eta}_{\textrm{TS}} (\tilde{t}) = \tilde{s} = N_0 - \tilde{l}$ together with the fact that $\tilde{l} = \tilde{R}$, 
one can obtain the final scaling form of the total translocation time as
\begin{equation}
\tilde{\tau} = \frac{1}{\tilde{f}} \big[ \tilde{\eta}_{\textrm{p}} N_0 + N_0^2 \big] .
\label{scaling_trans_time_Rod_like}
\end{equation}
Here the pore friction term, $\tilde{\eta}_{\textrm{p}} N_0 /\tilde{f}$, has a significant correction to asymptotic 
scaling similar to the flexible case and the effective exponents intermediate values
of $N_0$ will be between unity and two. We note that the scaling form of Eq. (52) has recently been
derived from an nonequilibrium transport theory
in the limit of drift-driven polymer translocation where the electrostatic interactions are weak \cite{poly2018}.

\section{Summary and conclusions}

In this article, we have presented a comparative review of the electrohydrodynamic and configurational approaches to polymer translocation. The distinction between these two approaches is based on the relative spatial scales of the polymer and the nanopore. The first part of the article is devoted to the translocation of short polymers whose size is comparable to the length of the translocated nanopore. At this scale, an accurate modeling of the translocation process requires the explicit consideration of the electrostatic and hydrodynamic details of the pore medium, such as the electrophoretic and hydrodynamic drag forces on the translocating molecule, and the electrostatic interactions of the molecule with the membrane and the surrounding electrolyte.

First, in Section~\ref{elh}, we have presented the electrohydrodynamic model of polymer translocation where these details are explicitly and consistently included via the coupled solution of the Stokes, Poisson-Boltzmann, and polymer diffusion equations. We then discussed the application of the model to various experimental configurations. In Section~\ref{resmf},  we presented direct comparisons with pressure-voltage-driven polymer trapping experiments carried out in monovalent salt solutions where the translocation process is governed by MF electrohydrodynamics. For this experimental setup, we showed that the electrohydrodynamic theory can quantitatively reproduce and explain the pressure dependence of the experimental polymer translocation velocity and time data. 

Then, in Section~\ref{pol}, we focused on polymer translocation experiments conducted in polyvalent electrolytes where the high ion valency drives the system out of the MF electrohydrodynamic regime. We showed that via the inclusion of charge correlations, the electrohydrodynamic theory can predict with quantitative accuracy the experimentally observed inversion of the electrohydrodynamic DNA mobility by added multivalent cations. We elucidated the electrohydrodynamic mechanism underlying the mobility reversal in terms of the DNA charge inversion. Finally, in Section~\ref{den}, we considered translocation experiments conducted with $\alpha$HL nanopores where the high pore confinement results in strong correlation effects even with monovalent electrolytes. Within the framework of the model, we showed that the experimentally observed salt-induced sharp rise and non-uniform voltage dependence of polymer capture rates can be explained by the competition between the electric drift force and surface polarization effects acting as a repulsive barrier for polymer capture.

Section~\ref{IFTP} was devoted to the opposite case of polymers much longer than the nanopore thickness. In this regime, as the pore electrohydrodynamics acts only on a small portion of the polymer, one can make the approximation to absorb the electrohydrodynamic details of the nanopore into an effective force $f$ acting locally on the polymer portion confined to the pore, and the effective (constant) pore friction $\eta_{\rm p}$. This simplification allows to bypass the electrohydrodynamic details of the translocation process, enabling accurate modeling of the configurational effects originating from conformational polymer fluctuations. Within the corresponding IFTP theory, we discussed the detailed characterization of the translocation dynamics of polymers with arbitrary length. We presented the predictions of the IFTP theory for the scaling of the polymer translocation time with the polymer length and the variation of the former with the pore friction and the external force driving the polymer. The theory is applicable to a variety of translocation and polymer pulling scenarios, including pore-driven and end-pulling setups discussed here, and is in excellent agreement with MD simulations of the corresponding coarse-grained polymer models. 

The grand challenge in polymer translocation consists of amalgamating the regimes of short and long polymers so far separately studied. This requires the consideration of electrohydrodynamic effects and conformational polymer fluctuations on the same footing. In our recent work of Ref.~\cite{poly2018}, a first attempt in this direction was made by incorporating the electrostatic coupling of the membrane with the \textit{cis} and \textit{trans} portions of the polymer outside the nanopore into the stiff polymer limit of the IFTP theory. We are currently working on the relaxation of the stiff polymer constraint. This extension will hopefully allow to better understand the effect of electrostatic interactions on the scaling of the translocation time with the polymer length.

\appendix

\section{Coefficients of the average velocity $\lan v_\p\ran$ in Eqs.~(\ref{vp2}) and~(\ref{bv})}
\label{ap1}

We report here the coefficients of the average polymer translocation velocity in Eqs.~(\ref{vp2}) and~(\ref{bv}).  
\bea
J_1&=&\frac{1}{\left(\ld-\lb\right)^2}\left\{\left(\ld-\lb\right)L_-+e^{-(\ld-\lb)L_-}-1\right\}\\
&&+\frac{1}{\ld-\lb}\left[1-e^{-(\ld-\lb)L_-}\right]\left\{\frac{1}{\ld}\left[1-e^{-\ld(L_+-L_-)}\right]+\frac{1}{\ld+\lb}e^{-\ld(L_+-L_-)}\left[1-e^{-(\ld+\lb)L_-}\right]\right\};\nonumber\\
J_2&=&\frac{1}{\ld^2}\left\{\ld(L_+-L_-)+e^{-\ld(L_+-L_-)}-1\right\}+\frac{1-e^{-(\ld+\lb)L_-}}{\ld(\ld+\lb)}\left[1-e^{-\ld(L_+-L_-)}\right];\nonumber\\
J_3&=&\frac{1}{\left(\ld+\lb\right)^2}\left\{\left(\ld+\lb\right)L_-+e^{-(\ld+\lb)L_-}-1\right\}.\nonumber\\
f_p&=&\mathrm{K}_1(\td)\mathrm{I}_0(\ta)+\mathrm{I}_1(\td)\mathrm{K}_0(\ta)-\left(\td\right)^{-1};\\
f_m&=&\mathrm{K}_1(\ta)\mathrm{I}_0(\td)+\mathrm{I}_1(\ta)\mathrm{K}_0(\td)-\left(\ta\right)^{-1};\\
g&=&\mathrm{I}_1(\td)\mathrm{K}_1(\ta)-\mathrm{I}_1(\ta)\mathrm{K}_1(\td).
\eea

\section{Components of the polymer translocation time $\tau$ in Eq.~(\ref{tauc2})}
\label{ap2}

We provide here the components of the translocation time in Eq.~(\ref{tauc2}).
\bea
\label{38I}
\tau_1&=&\frac{1}{D(\ld-\lb)^2}\left[e^{-(\ld-\lb)\lm}-1+(\ld-\lb)\lm\right];\\
\label{38II}
\tau_2&=&\frac{1}{D\ld(\ld-\lb)}\left[1-e^{-(\ld-\lb)\lm}\right]\left[1-e^{-\ld(\lp-\lm)}\right]+\frac{1}{D\ld^2}\left[e^{-\ld(\lp-\lm)}-1+\ld(\lp-\lm)\right];\nonumber\\
&&\\
\label{38III}
\tau_3&=&\frac{1}{D(\ld+\lb)^2}\left[e^{-(\ld+\lb)\lm}-1+(\ld+\lb)\lm\right]\\
&&+\frac{e^{-\ld(\lp-\lm)}}{D(\ld+\lb)}\left[1-e^{-(\ld+\lb)\lm}\right]\left\{\frac{1}{\ld-\lb}\left[1-e^{-(\ld-\lb)\lm}\right]+\frac{1}{\ld}\left[e^{\ld(\lp-\lm)}-1\right]\right\}.\nonumber
\eea

\end{document}